\RequirePackage{fix-cm}
\documentclass[twocolumn,epjc3]{svjour3}  
\RequirePackage[numbers,sort&compress]{natbib}

\smartqed  % flush right qed marks, e.g. at end of proof
\usepackage{amsmath,amssymb}
\usepackage{upgreek}
\usepackage{graphicx}
\graphicspath{{Figures/}}
\usepackage{latexsym}
\usepackage{url,hyperref}
\usepackage{bm}
\usepackage{textcomp}
\usepackage[dvipsnames]{xcolor} %[expansion]
\usepackage{bbm}
\usepackage{slashed}
\usepackage{epstopdf}
\usepackage{placeins}
\usepackage{array}
\usepackage{cancel}
\usepackage{changes}
\usepackage{verbatim}
\usepackage{csquotes} %for enquote

\usepackage{outlines} %nested lists! %\1 \2 \3
\usepackage{tensor} %neat indices

\usepackage{blindtext}
\usepackage{tcolorbox}

\newcommand{\beq}{\begin{equation}}
\newcommand{\eeq}{\end{equation}}

\hypersetup{colorlinks=true,linkcolor=blue,citecolor=magenta,filecolor=magenta,urlcolor=blue}

\DeclareUnicodeCharacter{2212}{-} %for \maketitle

\usepackage{etoolbox}

\makeatletter
\patchcmd{\frontmatter@abstract@produce}
  {\vskip200\p@\@plus1fil
   \penalty-200\relax
   \vskip-200\p@\@plus-1fil}
  {}
  {}
  {}
\makeatother
\newcolumntype{C}{>{$}c<{$}} % math-mode version of "l" column type
\newcolumntype{L}{>{$}l<{$}} % math-mode version of "l" column type

\journalname{Eur. Phys. J. C}
\begin{document}

\title{Black holes and nilmanifolds: quasinormal modes as the fingerprints of extra dimensions?%\thanksref{t1}
}

%\titlerunning{Short form of title}        % if too long for running head

\author{Anna Chrysostomou\thanksref{AC,UJ,IP2I}
        \and
        Alan Cornell\thanksref{ASC,UJ,IP2I} 
        \and
        Aldo Deandrea\thanksref{AD,UJ,IP2I}
        \and
        {\'E}tienne~Ligout\thanksref{EL,ENS}
        \and
        Dimitrios Tsimpis\thanksref{DT,IP2I}
}

%\thankstext{t1}{Grants or other notes
%about the article that should go on the front page should be
%placed here. General acknowledgments should be placed at the end of the article.
\thankstext{AC}{e-mail: \url{chrysostomou@ip2i.in2p3.fr} (corr. author)}
\thankstext{ASC}{e-mail: \url{acornell@uj.ac.za}}
\thankstext{AD}{e-mail: \url{deandrea@ip2i.in2p3.fr}}
\thankstext{EL}{e-mail: \url{etienne.ligout@etu.univ-paris1.fr}}
\thankstext{DT}{e-mail: \url{tsimpis@ip2i.in2p3.fr}}

%\authorrunning{Short form of author list} % if too long for running head

\institute{Department of Physics, University of Johannesburg, PO Box 524, Auckland Park 2006, South Africa \label{UJ}
           \and
           Institut de Physique des Deux Infinis de Lyon. Universit{\'e} de Lyon, UCBL, UMR 5822, CNRS/IN2P3. 4 rue Enrico Fermi, 69622 Villeurbanne Cedex, France. \label{IP2I}
           \and
            {\'E}cole Normale Sup{\'e}rieure de Lyon, 15 parvis Ren{\'e} Descartes, BP 7000, 69342 Lyon Cedex 07, France.\label{ENS}
}

\date{Received: date / Accepted: date}
% The correct dates will be entered by the editor

\maketitle

\begin{abstract}
We investigate whether quasinormal modes (QNMs) can be used in the search for signatures of extra dimensions. To address a gap in the Beyond the Standard Model (BSM) literature, we focus here on higher dimensions characterised by negative Ricci curvature. As a first step, we consider a product space comprised of a four-dimensional Schwarzschild black hole space-time and a three-dimensional nilmanifold (twisted torus); we model the black hole perturbations as a scalar test field. We suggest that the extra-dimensional geometry can be stylised in the QNM effective potential as a squared mass-like term representing the Kaluza-Klein (KK) spectrum. We then compute the corresponding QNM spectrum using three different numerical methods, and determine a possible ``detectability bound" beyond which KK masses cannot be detected using QNMs. 
% \PACS{PACS code1 \and PACS code2 \and more}
% \subclass{MSC code1 \and MSC code2 \and more}
\end{abstract}

\section{Introduction}
\par Quasinormal modes (QNMs), the damped and discrete oscillations in space-time that emanate from a perturbed body as it returns to an equilibrium state \cite{Vishveshwara1970,Press1971}, have served for several decades as a theoretical means of studying $d$-dimensional black hole space-times as well as a testing ground for the development of numerical simulations and techniques. Quantum gravity conjectures, modified theories of gravity, stability analyses of naked singularities and novel space-times, the development of numerical relativity simulations, and the explorations of the gauge-gravity duality are but a few of the avenues of research made accessible by black hole QNMs (see Refs. \cite{refNollert1999,refKokkotasRev,refFerrari2008,refBertiCardoso,refKonoplyaZhidenkoReview} for reviews). 

\par From the astrophysical perspective, quasinormal frequencies (QNFs) $\omega$ lead directly to insights about the nature of their black hole source. Specifically, the oscillation frequency $\mathbb{R}e \{ \omega \}$ and damping $\tau = -1/\mathbb{I}m \{\omega \}$ of the QNFs are uniquely determined by the characteristic black hole properties of mass $M$, spin $a$, and charge $Q$ \cite{Echeverria1989_BHpropertiesestimate}, in accordance with the \textit{no-hair conjecture} applicable to final-state black holes \cite{Gravitation1973}. This has earned QNMs the epithet ``black hole fingerprints" \cite{refKonoplyaZhidenkoReview}. As such, furthering our understanding of black holes in turn allows us to explore gravity in the relatively untested strong regime, complementing extant results obtained from experiments in low-velocity linear regimes \cite{Stairs2003_GRpulsarTests,Will2009_GRstellarTests,refWillReview2014,Berti2015_TestingGRastro}.

\par Mathematically, QNMs can be structured as an eigenvalue problem subjected to physically-motivated boundary conditions and dependent strictly on the features of the black hole space-time and the effective potential of the perturbing field. If we consider a static black hole (in an asymptotically-flat space-time) through the classical lens, radiation is purely in-going at the event horizon and purely out-going at spatial infinity; the black hole geometry is characterised solely by its mass \cite{NoHair_Schwarzschild} while the effective potential depends on the spin of the oscillating field and the multipolar (angular momentum) number $\ell$. For each $\ell$ there are infinitely many overtones $n$ labelling the QNF in increasing multiples of $\mathbb{I}m \{ \omega \},$ with the $n=0$ \textit{fundamental mode} representing the least-damped and thus longest-lived QNM. 

\par On the basis of spherical symmetry and time independence, the QNM behaviour in static black hole space-times can be shown to reduce to a simple radial wave equation, as first demonstrated in Refs. \cite{refRW,refZerilli} for the Schwarzschild case. A wide range of methods have been developed to determine QNM solutions from such wave equations, including methods that are ``exact" (e.g. direct integration methods \cite{DavisPrice1971,Price1994}, the continued fraction method \cite{refLeaver1985}, pseudospectral methods \cite{Dias2009_Paraspectral1,Dias2009_Paraspectral2}, etc.) and numerical (e.g. the asymptotic iteration method \cite{refAIM_OG,refAIM}, the Horowitz-Hubeny approach \cite{HorowitzHubeny}, etc.). Of these, we highlight \textit{(i) inverse-potential methods} that approximate the effective potential with an inverse P{\"o}schl-Teller potential \cite{PoschlTellerPotential} for which bound-state solutions are known to determine the QNF spectrum \cite{PoschlTellerMethod}; \textit{(ii) WKB-based methods} that adapt the semi-class- ical technique to the QNM problem to compute QNFs in the $\ell \geq n$ regime \cite{refBHWKB0,refBHWKB0.5,refBHWKB1} at sixth-order \cite{Konoplya2003} and beyond (see Ref. \cite{Konoplya2019}); \textit{(iii) photon-orbit methods} such as the inverse multipolar expansion method \cite{refDolanOttewill2009} that harnesses the known link between QNMs and unstable null geodesics \cite{refGoebel1972} to construct an iterative technique to solve the wave-like radial equation with increasing accuracy for large values of $\ell$ \cite{refOurLargeL}. This is by no means an exhaustive list; for further insights, we refer the reader to Refs. \cite{refBertiCardoso,refKonoplyaZhidenkoReview,Grandclement2007_Spectral,Dias2015_Num}.

\par Today, we find ourselves in the unprecedented position wherein which we can observe this gravitational radiation. To date, the LIGO-Virgo-KAGRA (LVK) collaboration has confirmed 90 gravitational-wave (GW) events with a probability of astrophysical origin\\
\noindent $p_{astro}~>~0.5$ \cite{refLIGO2018Run1,refLIGO2020Run2,refLIGO2021Run3a,refLIGOrecentRun3b}, providing us with the novel opportunity to scrutinise general relativity (GR) in the relativistic strong-field regime and placing us firmly in the era of GW astronomy. While this has immediate astrophysical \cite{Berti2015_TestingGRastro} and cosmological \cite{LVK2021_Cosmic} relevance, there is a significant interest in the theoretical implications of GWs \cite{Yunes2016_GW15implications} and the fundamental physics insights they might provide \cite{Barack2018_GWfunphys}.

\par This is in part due to their weakly-interacting nature: GWs propagate unimpeded through the universe, piercing both the cosmic microwave and cosmic neutrino backgrounds, possibly providing unique insights into the inflationary epoch and beyond \cite{Weir2017_GWsFOPTreview,Cai2017_GWphysics,Nature_GWreview}. These high and ultra high frequency {\textit{stochastic}} GWs correspond to energies of the TeV range and higher, towards the Planck scale; in this way, GWs serve as a complementary laboratory to collider physics experiments \cite{Aggarwal2020_HighFGWs,Alves_GWLHCComplement,Caldwell2022_GWLHCSnowMass}. Searches for new physics focused on early-universe dynamics are well-underway, with examples such as Ref. \cite{Huang2020_PatiSalamDynamics} demonstrating that models based at scales of Grand Unified Theories can be good candidates for detection via next-generation GW detectors \cite{refGWsensitivity}. This has encouraged new lines of inquiry into cosmic strings \cite{Chigusa2020_GWGUT,King2021_GWGUT}, leptogenesis \cite{White2019_GWGUT}, dark matter \cite{Buchmuller2019_GWGUT}, and other beyond the Standard Model (BSM) challenges.

\par GWs are also being applied to searches for extra dimensions (see Refs. \cite{Yu2019,Cardoso2019_GWsEDsKK,Kwon2019_GWsCompactifiedEDs}). Compact extra dimensions feature a variety of different geometries (Ricci-flat \cite{AndriotGomez2017_EDsignGWs,Du2021_GWsCompactifiedEDs,Ferko2022_GWsCompactifiedEDs}, toroidal \cite{AndriotTsimpis2019_WarpGWs}, warped toroidal \cite{AndriotTsimpisMarconnet2021_WarpGWs} extra dimensions, etc.). These, on the other hand, have so far predicted GWs whose frequencies are of the order of $10^{12}-10^{14}$ Hz, far exceeding the $10^3-10^4$ Hz upper limit of present and planned detectors \cite{Nature_GWreview,Aggarwal2020_HighFGWs}.

\par Here, however, we are guided by the capabilities of modern detectors, and investigate whether we can exploit present-day GW observations to infer constraints on new physics. In particular, we shall focus on binary black hole collisions, where the post-merger \textit{ringdown} phase is dominated by quasinormal ringing \cite{Berti2007}. For this reason, we can apply known theoretical and numerical QNM techniques to experimental observations. Logistically, we concentrate on black holes because the dynamics of binary black hole collisions have been studied extensively \cite{Berti2007,refFerrari2008,refBroeck2013}; the success of the LVK collaboration is a testament to the gravitational waveform modelling expertise, Bayesian statistical analysis techniques, and experimental prowess carefully honed over several decades (see Ref. \cite{refLIGOguide} for the LVK collaboration's guide on data acquisition and processing).

\par Furthermore, since the signal-to-noise ratio of the post-merger signal is usually fairly low and therefore not always characterisable \cite{LIGO2019_GWTC1-GRtest,LIGO2020_GWTC2-GRtest_pyRing3,LIGO2021_GWTC3-GRtest}, the higher-mass and louder black hole merger events are more likely to produce good candidates for ringdown analyses. The first detected GW event 
{\href{https://bilby-gwtc1.github.io/GW150914/html/bilby_bilby.html}{{\scshape{GW150914}}}} \cite{refLIGO,GW150914properties} was sufficiently loud to accommodate a QNM study, so we shall restrict our discussion to this event within this work, unless otherwise stated.

\par Current searches for evidence of new physics from available GW observations are dominated by model-agnostic null tests for deviations from GR predictions. These include: consistency checks between data and GR-based models for the evolution of a merger event; tests of the generation and propagation of GWs, where the latter involves searching for modifications to the dispersion relation and in turn constraining the Compton wavelength associated with the graviton mass; tests for additional polarisation modes beyond the tensor plus and cross modes predicted by GR; analyses of the post-merger properties for parametric deviations from GR \cite{LIGO2016_TestingGRGW150914,LIGO2019_GWTC1-GRtest,LIGO2020_GWTC2-GRtest_pyRing3,LIGO2021_GWTC3-GRtest}, etc. At present, there have been no statistically significant deviation from GR reported. 

\par However, this latter category of testing has been a subject of growing fascination, and motivates investment in more precise measurements of QNFs \cite{Berti2020_LISAringdownMultipolar,OtaChirenti2021_BHspectroscopyPresentFuture,Bhagwat2021_BHspectroscopyPresentFuture}. Furthermore, hopes for the establishment of \textit{black hole spectroscopy} \cite{Berti2005_BHspectroscopy} are beginning to be realised: although the $n=0$, $\ell=2$ mode is known to dominate the QNM spectrum, higher harmonics \cite{Carullo2019_pyRing1,Capano2021_MultipolarQNMsObserved} and overtones \cite{Cotesta2022_OvertonesRingdown} are being investigated. Tests of the no-hair conjecture are of particular interest \cite{Carullo2018_NoHairTests,refNoHair_pyRing,refNoHair_pyRing2,Isi2022_GW150914again,Ghosh:2021_QNMconstraints}, as a violation thereof may be evidence of an exotic object or new physics.

\par As such, we shall focus here on this use of parametric deviations from GR in the QNF spectrum in an attempt to outline a search for extra dimensions. In fact, there have already been considerations for extra dimensions using black hole QNMs, concentrated for example on the five-dimensional (5D) Randall-Sundrum II \cite{RSmodel2} model: through the formalism of Shiromizu {\textit{et al.}} \cite{Shiromizu1999_Einstein3Brane} and Dadhich {\textit{et al.}}, a 4D effective framework can be established from a 5D general relativity construction, leading to a (neutral) black hole solution that resembles the (charged) Reissner-Nordstr{\"o}m metric. The so-called {\textit{tidal charge}} $\beta= Q^2/(4M^2)$ is a manifestation of the influence of the extra dimension. In Ref. \cite{Mishra2021_QNMsEDs}, this is the observable utilised to constrain extra dimensions, but is found to disfavour extra dimensions.

\par It is not clear how to extend the Shiromizu {\textit{et al.}} formalism to a broader category of extra-dimensional models with $d>5$, nor is it obvious whether this tidal charge observable can be probed for general cases. Moreover, it may be that alternate geometries could lead to successful GW detection. Bearing these points in mind, we shall consider as an example a particularly simple partially-compactified setup: a direct product space featuring a four-dimensional Minkowski space-time and a three-dimensional negative compact space $\mathcal{M}_4 \times \mathcal{N}_3$. Within this space-time, we shall embed a four-dimensional Schwarzschild black hole. The higher-dimensional component will then be comprised of a twisted torus $-$ known as a \textit{nilmanifold} $-$ constructed from the non-trivial fibrations of layered tori. The nilmanifold is one of the few geometries that allows for analytic calculations of mass spectra and Kaluza-Klein (KK) reductions \cite{Andriot2016_TowardsKK}, and boasts a number of phenomen- ologically-interesting properties that we shall discuss in section \ref{sec:manifold}.

\par While the higher-dimensional manifold is highly specific, we shall show that the variable-separable nature of our extra-dimensional space-time in the absence of coupling between components of $\mathcal{M}_4$ and $\mathcal{N}_3$ and the QNM problem we consider, allows for a KK reduction that expresses the extra-dimensional behaviour as a mass-like term that can be incorporated into the QNM effective potential. We shall demonstrate how this enables the application of QNM literature on massive oscillating fields, as well as studies on parametric deviations from GR employed by the LVK collaboration. In so doing, we lay the groundwork for an additional avenue through which we may probe GW observations for model-agnostic extra-dimensional signatures. %This has the added benefit of relating QNM studies in the theoretical domain to observational data collected by experiments, which is not necessarily trivial.

\par The structure of this paper is as follows. In section \ref{sec:setup}, we define the ``Schwarzschild-nilmanifold" setup we investigate: we outline the interesting features and the construction of the nilmanifold, as well as the construction of the partially-compactified seven-dimensional (7D) metric and the scalar field\footnote{It is standard practice in QNM studies to explore uncharted space-times and/or novel techniques with scalar test fields to test for feasibility.} we use to explore it. There, we shall derive a 4D effective potential in which the higher-dimensional character is encoded in an effective mass term. In so doing, we recreate the problem of massive scalar QNMs: a scenario that has been used as a case study for numerical development in the QNM literature.

\par In section \ref{sec:QNFspectrum}, we compute the QNF spectrum using the three numerical methods highlighted, \textit{viz.} the inverse P{\"o}schl-Teller potential method, the WKB method, and the inverse-$\ell$ method. We include also a discussion on how the QNF spectrum is affected by the mass-like term. In this way, we shall determine an upper bound under which detectable black hole QNMs may serve as an appropriate probe for extra dimensions in this construction. To constrain this mass-like parameter further, we introduce bounds from studies on the parametric deviation of GW data from GR predictions, using the most stringent results published by the LVK collaboration \cite{LIGO2021_GWTC3-GRtest}. This step shall be carried out in section \ref{sec:GWs}. By comparing the magnitude of the deviation from GR in the ringdown phase with the deviations in a the QNF spectrum caused by our introduction of the effective mass term, we are able to place naïve constraints on detectable QNFs harbouring extra-dimensional signatures. In other words, we demonstrate a plausible detectability bound on the observation of KK masses using QNFs. While our interest lies specifically in the case of negative extra-dimensional components, this result is agnostic to the extra-dimensional scalar curvature and therefore applies to a wide variety of extra-dimensional setups featuring compact spaces. We note, however, that our objective here is \textit{not} to supply a definitive constraint on extra dimensions. Rather, we suggest a pragmatic method by which QNFs can be repurposed for BSM searches through combining known techniques and available GW data.

\section{A Schwarzschild-nilmanifold extra-dimensional setup \label{sec:setup}}
\par Compact negative-curvature spaces (i.e. spaces with negative Ricci scalar curvature) have been interrogated extensively within the mathematical literature \cite{Needham2012_VisualCA,Berger2013_PanoramicRiemannian}. Among members of the string theory community, a burgeoning interest in such spaces is developing in the wake of a recent observation that negatively-curved manifolds are a requirement for classical de Sitter solutions with orientifold planes \cite{Haque2008_NegCurve1,Andriot2016_NegCurve2,AndriotMarconnet2022_10dAdS}. In the context of particle physics, extra-dimensional models characterised by partial or total negative scalar curvature remain comparatively under-explored. 

\par Phenomenologically, studies on compact hyperbolic spaces are promising for their capacity to include cosmological observations such as homogeneity and flatness \cite{Starkman2001_CosmoED,Chen2003_CosmoHypED,Neupane2003_CosmoExpPot}. Moreover, these models could be used to address the hierarchy problem between the Planck and the electroweak scale by virtue of their geometrical properties. Compact negative-curvature spaces possess two characteristic length scales:  $\ell_{\rm c}$, associated with local properties like the curvature and fixed by the equations of motion, and $\ell_{\rm G}$, associated with global properties like the volume and independent of the equations of motion. Their volume grows exponentially with $\ell_{\rm c}/\ell_{\rm G}$, leading to an exponential reduction of the Planck length, which in turn yields a natural explanation for the perceived discrepancy in energy scales \cite{OrlandoPark2010_HyperbolicEDatLHC}. Furthermore, the KK mass spectra associated with such spaces are usually similar to those of Randall-Sundrum models \cite{RSmodel1} in that they accommodate the electroweak-Planck scale hierarchy without introducing light KK modes \cite{OrlandoPark2010_HyperbolicEDatLHC}.

\par Motivated by these implications, a series of investigations \cite{Andriot2016_TowardsKK,
AndriotTsimpis2018_Laplacian,
IP2I2020_Nilmanifolds,
Deandrea2022_GHnilmanifolds,
Deandrea2022_Diracnilmanifolds} have focused model-building efforts on a compact, negatively-curved manifold whose tangent vectors form a Lie algebra that is nilpotent \textit{viz.} a nilmanifold $\mathcal{N}_3$ (see Refs. \cite{Andriot2015_Solvmanifolds,Bock2009_Solvmanifolds}). In the sections that follow, we shall outline how the nilmanifold is constructed from the Heisenberg algebra and demonstrate the KK expansion of a scalar field in this context, as established in Ref. \cite{Andriot2016_TowardsKK}. With these elements in place, we may proceed to the construction of our Schwarzschild-nilmanifold setup, and the KK reduction that allows us to treat the oscillations travelling through the 7D product space-time as a massive 4D scalar field.

\subsection{Algebra, geometry, and a 3D scalar field \label{sec:manifold}}

\par Any Lie group of dimension $d$ can be understood as a $d$-dimensional differentiable manifold. Under certain conditions (see Ref. \cite{Grana2006_TwistedToriReview} for a review), a solvable\footnote{A Lie group $G$ is solvable if its Lie algebra $\mathfrak{g}$ terminates in the null algebra i.e. the sequence $\mathfrak{g}_0=\mathfrak{g}$, $\mathfrak{g}_{n+1}=[\mathfrak{g}_n,\mathfrak{g}_n]$ for $n\geq 0$ reduces to the null algebra after a finite number of steps.} Lie group $G$ can be divided by a lattice $\Gamma$, a discrete subgroup of $G$, to construct a compact solvmanifold (i.e. a twisted torus) by means of discrete identifications \cite{Andriot2010_Thesis}. %The dimension of the resultant quotient space $G/\Gamma$ is the same as that of the group $G$ \cite{Andriot2010_Thesis}.
\textit{Nilpotent}\footnote{A Lie group $G$ is nilpotent if the sequence $\mathfrak{g}_{n+1}=[\mathfrak{g},\mathfrak{g}_n]$ reduces to the null algebra after a finite number of steps.} groups are a special subclass of solvable groups. For them, the compactness criterion  requires the structure constants to be rational in some basis \cite{Malcev1951}. We refer to their corresponding compact manifolds as \textit{nilmanifolds}.
%A group for which such a $\Gamma$ always exists is called \textit{nilpotent}; the quotient manifold $G/\Gamma$ is a compact \textit{nilmanifold} \cite{Malcev1951}.  

\par Consider the $d$-dimensional Lie algebra $\mathfrak{g}$ generated by the vectors $\{ Z_a, a = 1, ..., d\}$ satisfying
\begin{equation}
[ Z_b, Z_c ] = f\indices{^a _{bc}}Z_a \;.
\end{equation} 
\noindent Here, the structure constants satisfy $f\indices{^a _{bc}} = -f\indices{^a _{cb}}$. The corresponding $d$-dimensional manifold admits a globally-defined orthonormal frame $ \{ e^a, a = 1,...,d \}$ (where this basis defines the dual space of one-forms $\mathfrak{g}^{\star}$). This frame obeys the Maurer-Cartan equation
\begin{equation} \label{eq:MaurerCartan}
\text{d}e^a= - \frac{1}{2} f\indices{^a _{bc}} e^b \wedge e^c = -\sum_{b<c}f\indices{^a _{bc}} e^b \wedge e^c \;,
\end{equation}
\noindent with the exterior derivative $\text{d}$. Since the dual space $\mathfrak{g}^{\star} \approx T_e G^{\star}$, $\{ e^a, a = 1,...,d \} $ provides $-$ by left invariance $-$ a basis for the cotangent space $T_x G^{\star}$ at every point $x \in G$, the one-forms are globally defined on the manifold. These one-forms will have their non-trivial identification through the ``lattice action" when $G$ is divided by $\Gamma$. Note that $ f\indices{^a _{bc}}$ is related to the spin connection.

\par In flat indices and for a unimodular Lie algebra, the Ricci tensor is given by
\begin{eqnarray}
\mathcal{R}_{cd} &=& \frac{1}{2} \Big(- f\indices{^b _{ac}} f\indices{^a _{bd}} - \delta^{bg} \delta_{ah} f\indices{^h_{gc}} f\indices{^a _{bd}} \nonumber \\
&& +\frac{1}{2} \delta^{ah} \delta^{bj} \delta_{ci} \delta_{dg} f\indices{^i _{aj}} f\indices{^g _{hb}} \Big) \;,
\end{eqnarray}
\noindent with $\delta_{ab}$ serving as a Euclidean metric. For the nilpotent algebra, and thus for the nilmanifold case, the first term vanishes. The Ricci tensor is thus nowhere-vanishing and the corresponding Ricci scalar emerges as
\begin{equation}
\mathcal{R} = - \frac{1}{4} \delta_{ad} \delta^{be} \delta^{cg} f\indices{^a _{bc}} f\indices{^d _{eg}} \;.
\end{equation}
\noindent The Ricci scalar is strictly negative.

\par From Eq. (\ref{eq:MaurerCartan}), we can see that $d=3$ is the lowest dimensionality for which this expression is non-trivially satisfied. %From the phenomenological standpoint, this suits us: if $d \geq 3$, Mostow's rigidity theorem implies that geometrical quantities such as the length of the geodesics or the volume are topological invariants so that there is no additional moduli once the geometrical parameters $\ell_{\rm c}$ and $\ell_{\Gamma}$ are fixed \cite{Mostow1968_Rigidity}. [This applies only to hyperbolic manifolds -- it does not pertain to nilmanifolds!]
For $d=3$, there is the trivial Abelian algebra that leads to a three-torus, as well as three different solvable algebras. Of these, one is nilpotent: the \textit{Heisenberg} algebra 
\begin{equation} \label{eq:Heisenberg}
[ Z_1, Z_2 ] = - \mathtt{f}Z_3 \;, \hspace{0.3cm} [Z_1,Z_3 ] =[ Z_2, Z_3 ] = 0 \;, 
\end{equation}
\noindent with $\mathtt{f} = -f^3_{12} \neq 0$ such that the Maurer-Cartan equation becomes
\begin{equation} \label{eq:MC}
\text{d}e^3 = \mathtt{f} e^1 \wedge e^2 \;, \hspace{0.3cm} \text{d}e^1 = 0 \;, \hspace{0.3cm} \text{d}e^2 = 0 \;.
\end{equation}
\noindent The only nonzero structure constant $\mathtt{f}=-f^3_{\;\;\;12} \in \mathbb{R}$ is the \textit{geometric flux} serving as the nilmanifold's \textit{twist parameter}. The corresponding geometric properties of the nilmanifold can be relayed through the Maurer-Cartan equation Eq. (\ref{eq:MC}), from which we define
\begin{equation} \label{eq:radii}
e^1 = r^1 \mathrm{d}y^1 \;, \; \; e^2 = r^2 \mathrm{d}y^2 \;, \;\; e^3 = r^3 (\mathrm{d}y^3 + N y^1 \mathrm{d}y^2 )
\end{equation}
\noindent for the constant radii $r^{1,2,3}>0$, angular coordinates $y^m \in [0,1]$, and the integer $N = r^1 r^2 \mathtt{f} / r^3$ \cite{Andriot2016_TowardsKK}.   

\par The discrete identifications that make the compactification possible are
\begin{equation} \label{eq:identifications}
    y^1 \sim y^1 + n^1 \;, \;\; y^2 \sim y^2 + n^2 \;, \;\; y^3 \sim y^3 + n^3 - n^1 N y^2 \;,
\end{equation}
\noindent for $n^{m=1,2,3} \in [0,1].$ In other words, these identifications correspond to the lattice action responsible for establishing $\mathcal{N}_3$ as a nilmanifold. Eq. (\ref{eq:identifications}) leaves Eq. (\ref{eq:radii}) invariant.
\par In this way, the compact manifold is fully characterised as a twisted $S^1$ fibration over layered tori $T^2$. The twist is along the fibre coordinate $y^3$, while the base is parameterised by the coordinates $(y^1,y^2)$. Physically, $y^{m=1,2,3}$ are angles defined on $[0,1]$. The constant radii $r^m$ have units of length, the coordinates $y^m$ are dimensionless, and $\mathtt{f}$ has units of inverse length (i.e. energy). 

\par The most general left-invariant metric for the nilmanifold is given by
\begin{equation}
    \text{d}s^2 = \delta_{ab} E^a E^b \;, \;\;\; E^a = \left(L^{-1} \right)\indices{^a _{b}}e^b , 
\end{equation}
\noindent where we use $E^a$ to denote the one-forms related to the orthonormal basis $e^a$ through the constant $GL(3,\mathbb{R})$ transformation $L$.

\par To demonstrate the construction of the scalar mass spectrum, we shall consider the simplified special case in which $r^m=1$ and $\mathtt{f}=1$. The nilmanifold metric then becomes
\begin{equation} \label{eq:nil}
    \mathrm{d}s^2_{\rm nil} = \delta_{ab} e^{a} e^{b} =  (\mathrm{d}y^1)^2 + (\mathrm{d}y^2)^2 + ( \mathrm{d}y^3 + y^1 \mathrm{d}y^2 )^2 \;.
\end{equation}
\noindent 
\par To understand the behaviour of a scalar field on this space, we consider the massive Klein-Gordon equation. Let us begin with the Laplacian
\begin{equation}
    \nabla^2 \Phi = \frac{1}{\sqrt{g}} \delta_m \left(\sqrt{g} g^{mn} \delta_n \Phi \right) \;,
\end{equation}
\noindent where the determinant $\sqrt{g}=r^1r^2r^2$ reduces to 1 in our simplified metric. We may write
\begin{equation} \label{eq:LaplaceU}
    \nabla^2 u = \left( \partial_1^2 + \left(\partial_2 - y^1 \partial_3 \right)^2 + \partial_3^2 \right) u \;, 
\end{equation}
\noindent as we shall consider the expansion of $u$ on the space of functions invariant under Eq. (\ref{eq:identifications}), beginning with the functions depending only on the base coordinates $(y^1,y^2)$. In this case, the Laplacian is easily diagonalised:
\begin{equation}
    \left( \nabla^2 + \upmu^2_{\beta,\gamma} \right) {\tilde{v}}_{\beta,\gamma} = 0 \;, 
\end{equation}
\noindent where we define
\begin{equation}
\tilde{v}_{\beta,\gamma} (y^1,y^2) = e^{2 \pi i \beta y^1}  e^{2 \pi i \gamma y^2} \;, 
\end{equation}
for $\beta,\gamma \in \mathbb{Z}$, as invariant under Eq. (\ref{eq:identifications}), and the Klein-Gordon masses as
\begin{equation} \label{eq:basemass}
    \upmu^2_{\beta,\gamma}  =  4 \pi^2 \left( \beta^2 + \gamma^2 \right) \;.
\end{equation}

\par We can present a more generalised expression using the \textit{Weil-Brezin-Zak} transforms \cite{Thangavelu_HA} for a basis of invariant functions $u_{\kappa,\lambda}$,
\begin{eqnarray} \label{eq:WBZ}
    u_{\kappa,\lambda} (y^1,y^2,y^3) &=& e^{2 \pi \kappa i (y^3 + y^1y^2)} e^{2 \pi \lambda i y^1} \\
&& \times \sum_{\sigma} e^{2 \pi \kappa \sigma i y^1} f(y^2 + \sigma) \;,\nonumber
\end{eqnarray}
for $\kappa, \lambda, \sigma \in \mathbb{Z}$. Since $u_{\kappa,\lambda}$ is invariant under Eq. (\ref{eq:identifications}) for all values of $f(x)$, the functions remain well-defined across our nilmanifold $\mathcal{N}_3$. Upon substituting Eq. (\ref{eq:WBZ}) into Eq. (\ref{eq:LaplaceU}), we obtain
\begin{eqnarray}
\nabla^2 u_{\kappa,\lambda}  &=& e^{2 \pi \kappa i (y^3 + y^1y^2)} e^{2 \pi \lambda i y^1} \sum_{\sigma} e^{2 \pi \kappa \sigma i y^1} \\
& \times& \left[ \partial_2^2 - 4 \pi^2 \left( \kappa^2 + (\kappa (y^2 + \sigma) + \lambda)^2 \right) \right] f(y^2 + \sigma ) \;, \nonumber
\end{eqnarray}
\noindent where we require that $\kappa \neq 0$ to retain the $y^3$-dependent terms. 

\par If we introduce $z_{\sigma} = y^2 + \sigma + \lambda/\kappa$ and $g(z_{\sigma} ) = f(y^2 + \sigma)$, we can rewrite the above Laplacian as

\begin{eqnarray} \label{eq:LaplaceUz}
\nabla^2 u_{\kappa,\lambda}  &=& e^{2 \pi \kappa i (y^3 + y^1y^2)} e^{2 \pi \lambda i y^1} \sum_{\sigma} e^{2 \pi \kappa \sigma i y^1} \\
& \times& \left[ \partial_{z_{\sigma}}^2 - (2 \pi \kappa)^2 \left( z_{\sigma}^2 + 1) \right] g(z_{\sigma}) \right]\;. \nonumber
\end{eqnarray}
\par From the normalised Hermite functions 
\begin{equation} 
X_{\nu} (z) = e^{-z^2/2} H_{\nu} (z) \;, \;\; \nu \in \mathbb{N} \;,\end{equation}
where $H_{\nu}$ represents the Hermite polynomials, we may define
\begin{equation} \label{eq:HermiteX}
X_{\nu}^{\rho} (z) = \vert \rho \vert^{1/4} X_{\nu} ( \vert \rho \vert^{1/2} z) 
\end{equation}
\noindent for $\rho \in \mathbb{R}^*$ \cite{Thangavelu_HA}. By the properties of Hermite polynomials, Eq. (\ref{eq:HermiteX}) satisfies the differential equation
\begin{equation}
   ( \partial^2_z - \rho^2 z^2 ) X^{\rho}_{\nu} (z) = - (2\nu +1) \vert \rho \vert X^{\rho}_{\nu} (z) \;.
\end{equation}
\par With the insertion of $g(z_{\sigma}) = X^{2 \pi \kappa}_{\nu} (z_{\sigma})$ into Eq. (\ref{eq:LaplaceUz}), we obtain the 3D Klein-Gordon equation
\begin{equation}
    \left( \nabla^2 + M^2_{\kappa,\lambda,\nu} \right) \tilde{u}_{\kappa,\lambda,\nu} = 0 \;,
\end{equation}
\noindent where the masses and wavefunctions are, respectively,
\begin{equation}
M^2_{\kappa,\lambda,\nu} = (2 \pi \kappa)^2 \left( 1+\frac{2 \nu + 1}{2 \pi \vert \kappa \vert} \right) \;, \label{eq:fibremass}
\end{equation}
\begin{eqnarray}
    \tilde{u}_{\kappa,\lambda,\nu} (y^1,y^2,y^3) & = & e^{2 \pi \kappa i (y^3 + y^1y^2)} e^{2 \pi \lambda i y^1} \sum_{\sigma} e^{2 \pi \kappa \sigma i y^1} \nonumber \\
&& \times  X^{2 \pi \kappa }_{\nu} \left(y^2 + \sigma + \frac{\lambda}{\kappa} \right) 
\end{eqnarray}
for $\sigma \in \mathbb{Z}$, $\nu \in \mathbb{N}$, $\kappa \in \mathbb{Z}^*$, and $\lambda = 0, ... , \vert \kappa \vert -1 $. The range of $\lambda$ is derived from the fact that $\lambda$ itself is defined modulo $\kappa$, which in turn is a consequence of the identity
\begin{equation}
    \tilde{u}_{\kappa, \lambda + \kappa \tau, \nu} (y^1,y^2,y^3) = \tilde{u}_{\kappa, \lambda , \nu} (y^1,y^2,y^3) \;\; \forall \; \tau \in \mathbb{Z} \;.
\end{equation}
\par By virtue of Eq. (\ref{eq:fibremass})'s independence of $\lambda$, there exists a mass degeneracy. The wavefunctions are parameterised by a finite number of inequivalent values of $\lambda$ such that the level of the degeneracy is $\vert \kappa \vert.$ Note that only one zero-mode (i.e. with vanishing mass) exists for this Klein-Gordon equation, $\tilde{v}_{0,0}$, corresponding to the modes of the torus base.

\par We conclude this discussion on the nilmanifold space with the physical spectrum associated with a scalar field propagating on $\mathcal{N}_3$. This is achieved by reintroducing dimensional parameters $r^m$ and $\mathtt{f}$ \cite{Andriot2016_TowardsKK}. We may distinguish between \textit{torus modes},
\begin{eqnarray}
v_{\beta,\gamma} (y^1,y^2) & = & \frac{1}{\sqrt{V}} e^{2 \pi i \beta y^1} e^{2 \pi i \gamma y^2} \;,  \label{eq:torus_modes} \\
\upmu^2_{\beta,\gamma} & = & \beta^2 \left( \frac{2 \pi}{r^1} \right)^2 + \gamma^2 \left( \frac{2 \pi}{r^2}\right) \;, \label{eq:torus_mass}
\end{eqnarray}
and \textit{fibre modes},
\begin{eqnarray} 
    u_{\kappa,\lambda,\nu} (y^1,y^2,y^3) & = & \sqrt{\frac{r^2}{\vert N \vert V}} \frac{1}{\sqrt{2^{\nu} \nu! \sqrt{\pi}}}  e^{2 \pi \kappa i (y^3 + Ny^1y^2)}  \nonumber \\
& \times & e^{2 \pi \lambda i y^1}  \sum_{\sigma} e^{2 \pi \kappa \sigma i y^1} X^{\rho }_{\nu} (w_{\sigma}) \label{eq:fibre_modes} \;, \\
M^2_{\kappa,\lambda,\nu} &=&  \kappa^2 \left( \frac{2 \pi}{r^3} \right)^2 + (2 \nu + 1) \vert \kappa \vert \frac{2 \pi \mathtt{f}}{r^3}\;, \label{eq:fibre_mass}
\end{eqnarray}
\noindent for which we define
\[ \rho = \frac{2 \pi \mathtt{f}}{r^3} \kappa \;, \hspace{0.3cm} w_{\sigma}= r^2 \left( y^2 + \frac{\sigma}{N} + \frac{\lambda}{N\kappa} \right) \;,\]
and the volume
\begin{equation}
    V = \int d^3 y\, \sqrt{g} = r^1r^2r^3 \;.
\end{equation}

\par The scalar spectrum on the nilmanifold contains a complete
tower of modes on the torus that is independent of the fibre coordinate and radius. The fibre modes, whose mass spectrum is a function of the radial components and the curvature-related energy scale $\mathtt{f}$, have been shown to be tunable in Ref. \cite{Andriot2016_TowardsKK} by varying parameters in the generalised case; the fibre modes can be made lighter than their toroidal counterparts and the energy gaps in the spectrum may be enhanced. From the structure of Eq. (\ref{eq:fibre_mass}) itself, we understand that the fibre modes present with a unique mass spectrum: added to the typical $1/R$ Kaluza-Klein term is the novel $\mathtt{f}$-dependent term that enforces more finely-spaced modes, which follow a linear Regge trajectory. From the characteristic fibre-mode spectrum, we would expect a unique experimental signature.

{\color{black}{

\par To see clearly the distinctive spectrum of the nilmanifold, let us compare the fibre-mode masses of Eq. (\ref{eq:fibre_mass}) to 
the KK masses of a standard compactification $M_{st;\kappa,\lambda,\nu}$ on a three-dimensional torus $\mathbb{T}^3$, 
\begin{equation} 
 M^2_{st;\kappa,\lambda,\nu}= 
  \kappa^2 \left( \frac{2 \pi}{r^1} \right)^2 +
   \lambda^2 \left( \frac{2 \pi}{r^2} \right)^2 +
 \nu^2 \left( \frac{2 \pi}{r^3} \right)^2 
 ~,
\end{equation}
\noindent where $\kappa$, $\lambda$, $\nu\in\mathbb{Z}$. 
For simplicity we shall take all internal radii to be equal, $r^1=r^2=r^3$. Moreover we shall 
consider a nilmanifold $\mathcal{N}_3$ with minimal twist, $N=1$. 
The ratio $R$ of excited  KK masses to the lowest-lying one is then independent of 
the size of the radii of the internal manifold.  
For $\mathcal{N}_3$, $R^2_{nil}$ is given by 
\begin{equation} \label{eq:Rnil}
 R^2_{nil}= \frac{M^2_{\kappa,\lambda,\nu}}{M^2_{1,0,0}}=
 \frac{2\pi \kappa^2+(2\nu+1)|\kappa|}{1+2\pi}~.
\end{equation}
In contrast, for the standard $\mathbb{T}^3$,  $R^2_{st}$ is given by
\begin{equation} \label{eq:Rst}
 R^2_{st}= \frac{M^2_{st;\kappa,\lambda,\nu}}{M^2_{st;1,0,0}}=
\kappa^2+\lambda^2+\nu^2 ~.
\end{equation}
{
\begin{table}[h]
\addtolength{\tabcolsep}{-2pt}
  \begin{center}
\caption{\textit{The first few mass ratios $R_{nil}$ for the 3D nilmanifold $\mathcal{N}^3$, corresponding to  Eq.~(\ref{eq:Rnil}) for $\kappa=1$.}}\label{tab:specd=k1}
   \def\arraystretch{1.2}%
  \begin{tabular}{@{}CCCCCCCCCCCC@{}}
\hline\noalign{\smallskip}
  \nu  & 0 & 1 & 2 & 3 & 4 & 5 & 6 & 7& 8&9&10 \\
\hline\noalign{\smallskip}
     R^2_{nil}& 1.0 & 1.3 & 1.6 & 1.8 & 2.1 & 2.4 & 2.6 & 2.9 & 3.2 &3.5&3.7 \\
\hline\noalign{\smallskip}
    \end{tabular}
  \end{center}
\end{table}
\begin{table}[h!]
\addtolength{\tabcolsep}{-2pt}
  \begin{center}
\caption{\textit{The first few mass ratios $R^2_{nil}$ for the 3D nilmanifold $\mathcal{N}^3$, corresponding to  Eq.~(\ref{eq:Rnil}) for $\kappa=2$.}}\label{tab:specd=k2}
   \def\arraystretch{1.2}%
  \begin{tabular}{@{}CCCCCCCCCCCC@{}}
\hline\noalign{\smallskip}
  \nu  & 0 & 1 & 2 & 3 & 4 & 5 & 6 & 7& 8&9&10 \\
\hline\noalign{\smallskip}
     R^2_{nil}& 3.7 & 4.3 & 4.8 & 5.4 & 5.9 & 6.5 & 7.0 & 7.6 & 8.1 & 8.7 & 9.2 \\
\hline\noalign{\smallskip}
    \end{tabular}
  \end{center}
\end{table}

\begin{figure}[h] 
\begin{center}
\includegraphics[width=0.4\textwidth]{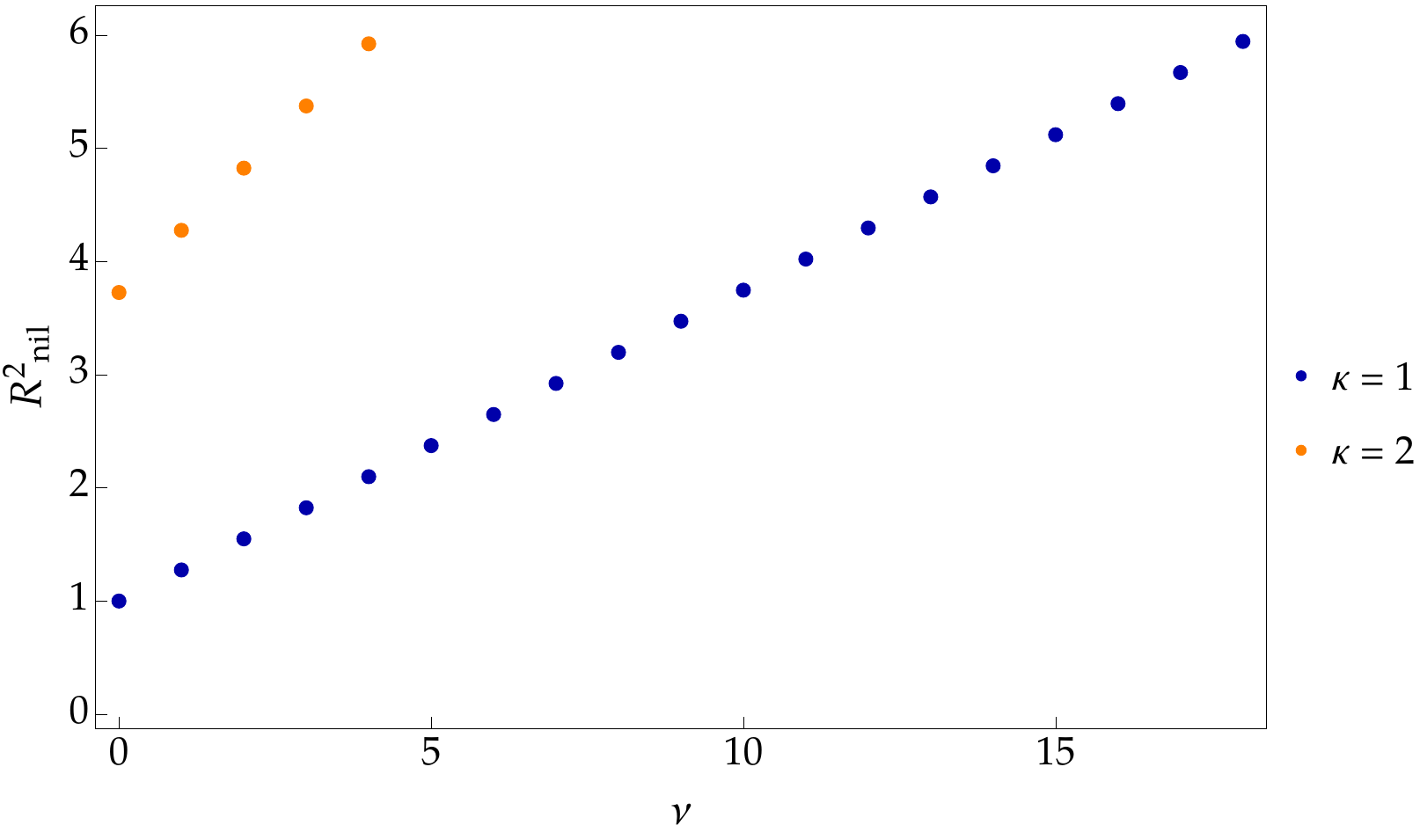}
\caption{\textit{Plot of the mass ratios $R^2_{nil}\leq6$ corresponding to  Eq.~(\ref{eq:Rnil}) for $\mathcal{N}_3$ for $\kappa=1$ (blue) and $\kappa=2$ (orange). 
}}\label{fig:1r}
\end{center}
\end{figure}

\begin{table}[ht]
  \begin{center}
  \caption{\textit{The first few mass ratios $R^2_{st}$ for the standard 3D torus $\mathbb{T}^3$, corresponding to Eq. (\ref{eq:Rst}).}}\label{tab:specd=2}
  \def\arraystretch{1.2}%
  \begin{tabular}{@{}CCCCCCC@{}}
\hline\noalign{\smallskip}
   (\kappa\lambda\nu) & (100) & (110) &(111) &(200)&(210)&(211) \\
\hline\noalign{\smallskip}
     R^2_{st} & 1 & 2 & 3 & 4 & 5 & 6 \\
\hline\noalign{\smallskip}
    \end{tabular}
  \end{center}
\end{table}
\begin{figure}[ht!]
\begin{center}
\includegraphics[width=0.38\textwidth]{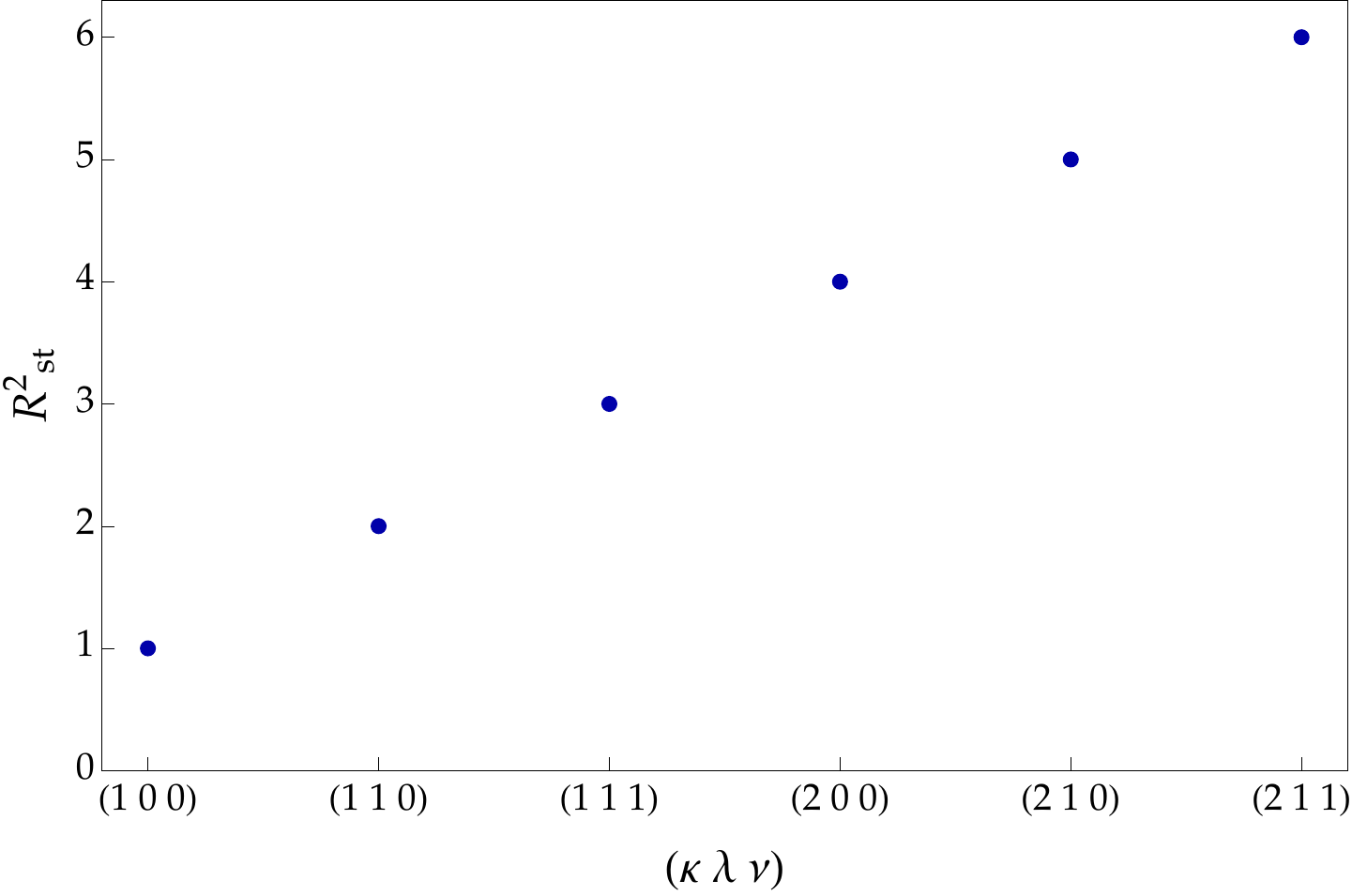} 
\caption{\textit{Plot of the ratios $R^2_{st}\leq6$ corresponding to  Eq.~(\ref{eq:Rst}) and Table \ref{tab:specd=2} for $\mathbb{T}^3$.}}\label{fig:2r}
\end{center}
\end{figure}
}
\par Since nilmanifolds allow for the possibility of analytically calculating the spectrum of propagating fields, they can be promising tools in the construction of effective BSM frameworks. Such models may be embeddable in string theory compactifications \cite{Andriot2016_TowardsKK}. As mentioned in the introduction, recent investigations into GW signatures of compact extra dimensions predict observables at frequencies of the order of $10^{12}-10^{14}$ Hz and higher \cite{AndriotGomez2017_EDsignGWs,AndriotTsimpis2019_WarpGWs,AndriotTsimpisMarconnet2021_WarpGWs,Yu2019,Cardoso2019_GWsEDsKK,Kwon2019_GWsCompactifiedEDs} $-$ several orders of magnitude beyond the $10^4$ Hz upper bound on modern detectors. However, these investigations suggest also that the KK GW spectrum is sensitive to changes in geometry. For example, introducing a non-trivial warp factor, as shown in Ref. \cite{AndriotTsimpisMarconnet2021_WarpGWs}, can lower the first KK mass by at least 69\% as compared against the standard KK spectrum on a torus $\mathbb{T}^d$. This is promising for the high-frequency GWs in extra-dimensional frameworks, as the relationship between frequency and KK mass implies that lower KK mass corresponds to GW frequencies closer to the sensitivity of modern instruments. 

\par In Figs. \ref{fig:1r} and \ref{fig:2r}, we see a similarly encouraging behaviour when we compare the fibre-mode spectrum with that of the standard torus modes. While we centre this work on the feasibility of detection with present-day data from the LVK collaboration, this effect motivates further investigation into the GWs propagating in nilmanifold spaces.
}}

\subsection{A Schwarzschild black hole and its scalar QNM}

\par GR remains our most complete theory of gravity to date. Its underlying principle is the relationship it defines between the geometry and matter content of a space-time, expressed concisely through the Einstein field equations,
\begin{equation} \label{eq:EFE}
    G_{MN} + \Lambda g_{MN} = \kappa_D\, T_{MN} \;.
\end{equation}
\noindent Here, the Einstein tensor $G_{MN}$ expresses the local space-time curvature, $\kappa_D$ is the Einstein gravitational constant in $D$ dimensions, and $T_{MN}$ is the stress-energy tensor that defines the energy, momentum, and stress for the matter and field content within the local space-time \cite{Gravitation1973}. We set $G=c=1$, unless otherwise stated. In asymptotically-flat space-times, the cosmological constant $\Lambda$ vanishes.  To describe the evolution of the metric and the fields, we utilise the $D$-dimensional Einstein-Hilbert gravitational action
\begin{equation} \label{eq:EHaction}
    \mathcal{S} = \frac{1}{2 \kappa_D} \int \mathrm{d}^{D} x\, \sqrt{-g}\, \left( R + \mathcal{L}_m \right)\;,
\end{equation}
\noindent where we use $\mathcal{L}_m$ to refer to all matter fields $\Psi$ within the space-time, and whose stress, energy, and momentum are encompassed by $T_{{MN}}$. 

\par Within the context of GR, Birkhoff's theorem stipulates that the most general spherically-symmetric vacuum solution of Eq. (\ref{eq:EFE}) is the Schwarzschild metric
\begin{eqnarray} \label{eq:Schwarz}
    \mathrm{d}s^2_{\rm BH} &=& g_{\mu \nu}^{\rm BH} \mathrm{d}x^{\mu} \mathrm{d}x^{\nu} = -f(r)\, dt^2 + f(r)^{-1}\, dr^2 \nonumber \\
    &&+ r^2(d \theta^2 + \sin^2 d\phi^2) \;,
\end{eqnarray}
\noindent where $f(r)=1-r_{H}/r$ and $r_{H}=2M$ is the Schwarzschild event horizon. For such a black hole, the length scale is defined by $M=mGc^{-2}$ for black hole mass $m^{\rm BH}$ \cite{Bekenstein2003_BHinfo}, and is set to unity. The Schwarzschild coordinates $(t,r,\theta,\phi)$ are defined on the regions $t \in (-\infty,+\infty)$, $r \in (r_H,+\infty)$, $\theta \in (0,\pi),$ and $\phi \in (0,2\pi)$; the tortoise coordinate $dr_*=dr/f(r)$ can be introduced to map the semi-infinite region of $(r_H,+ \infty)$ to $(-\infty,+\infty)$. 

\par Eq. (\ref{eq:Schwarz}) describes an isolated, static, and neutral 4D black hole \cite{Gravitation1973,Chandrasekhar1983} that is fully characterised by its mass $M$ \cite{NoHair_Schwarzschild}. Mathematically, black holes are therefore simple objects: they are pure geometry and do not require an equation of state to describe their evolution. Astrophysical black holes, however, are perpetually in a perturbed state: even if somehow isolated from the fields and matter in their immediate vicinity, they interact with the surrounding vacuum through Hawking radiation \cite{Hawking1975_HawkRad}. 

\par Black hole perturbation theory therefore considers a linearised approximation in which the black hole is described using
\begin{equation} \label{eq:pertmetric}
    g'_{\mu \nu} = g^{\rm BH}_{\mu \nu} + \delta_{\mu \nu} \;,
\end{equation}
\noindent where the unperturbed black hole metric $g^{\rm BH}_{\mu \nu}$ is referred to as the ``background" and the ``perturbations" $\delta_{\mu \nu}$ are considered to be very small ($\delta_{\mu \nu} \ll g^{\rm BH}_{\mu \nu}$). Similarly, we may consider a perturbed background field $\Psi'~=~\Psi^{\rm BG} + \psi$. We may then substitute $g'_{\mu \nu}$ and $\Psi'$ into Eq. (\ref{eq:EFE}), linearise the system of equations with respect to $\delta_{\mu \nu}$ and $\psi$, and thereby deduce the lineaised set of differential equations satisfied by the perturbations.  

\par As detailed in Chandrasekhar's book \cite{Chandrasekhar1983}, black hole QNM behaviour within a classical GR context can be inferred by substituting the perturbed metric, Eq. (\ref{eq:pertmetric}) and an ansatz into the Einstein field equations, and then solving for the vacuum solution under the physically-motivated QNM boundary conditions.\footnote{As stipulated in Ref. \cite{refKonoplyaZhidenkoReview}, at sufficiently late times, the QNMs obtained through this linear approximation remain in good agreement with those calculated via the full nonlinear integration of the Einstein equations \cite{refNonLinear1,refNonLinear2}.} The QNM ansatz and the number of ordinary differential equations required to describe the QNM propagation are derived from the symmetries of the background space-time: in the Schwarzschild case (static, non-rotating, and spherically-symmetric), the wave-function is written in variable-separable form,
\begin{equation} \label{eq:4Dphi}
    \Phi^s_{n \ell m} (\mathbf{x}) = \sum_{n}^{\infty} \sum_{\ell,m}^{\infty} \; \frac{\psi_{sn \ell}(r)}{r} \; Y^s_{m \ell} (\theta, \phi) \;,
\end{equation}
\noindent and the angular behaviour is relayed through spherical harmonics
\begin{equation}
    \label{eq:spherharmID} 
\nabla^2 Y^{s}_{m \ell} (\theta, \phi ) = -\frac{\ell (\ell + 1)}{r^2} Y^{s}_{m \ell} (\theta, \phi ) \;.
\end{equation}
\noindent Since the black hole is static, the corresponding ordinary differential equations are time independent. Consequently, the defining QNM behaviour is then fully encapsulated by the radial component.

\par As a simple example that retains the physical implications, we can consider Eq. (\ref{eq:4Dphi}) to be a scalar test field evolving on a fixed background in vacuum that contributes negligibly to the energy-density of the system. Explicitly, we may focus on the second term of Eq. ({\ref{eq:EHaction}}), which becomes
\begin{equation}
    \mathcal{L}_m = - \left( \partial_{\mu} \Psi \right)^{\dagger} \partial^{\mu} \Psi 
\end{equation}
\noindent for a minimally-coupled massless scalar field. The equations of motion satisfied by the fields $g_{\mu\nu}$ and $\Psi$ are then the massless Klein-Gordon equation for a curved space-time,
\begin{equation}
\nabla_{\mu} \nabla^{\mu} \Psi = \frac{1}{\sqrt{-g}} \partial_{\mu} \left( \sqrt{-g} g^{\mu \nu} \partial_{\nu} \Psi \right) = 0 
\end{equation}
and Eq. (\ref{eq:EFE}), with $T_{\mu\nu}$ quadratic in $\Psi$. In this context, the linearised equations of motion for $\psi$ and $\delta_{\mu \nu}$ decouple when $\Psi^{\rm BG}=0$, allowing for the metric fluctuations $\delta_{\mu \nu}$ to be set to zero. With the substitution of Eq. (\ref{eq:4Dphi}) into the above equation and the application of the tortoise coordinate, we obtain the radial wave-like equation sufficient to convey the QNM behaviour  

\begin{equation} \label{eq:SWEpsi}
    \frac{d^2\psi}{dr^2_*} + \left(\omega^2 - V(r) \right) \psi = 0 \;, 
\end{equation}
where
\begin{equation}
    V(r)=  f(r) \left(\frac{\ell (\ell+1)}{r^2}+\frac{f^{\prime}(r)}{r} \right) \;.
\end{equation}

\subsection{The effective 4D QNM problem}
\par In combining Eqs. (\ref{eq:Schwarz}) and (\ref{eq:nil}), we can construct our extra-dimensional manifold $ds_{7D}^2 = ds^2_{\rm BH} + ds^2_{\rm nil}$. In the absence of mixing terms, we consider a 7D scalar field propagating on this direct product space to be expressible as
\begin{equation}
 \Psi^s_{n \ell m} (\mathbf{z}) = \sum_{n=0}^{\infty} \sum_{\ell,m}^{\infty} \; \frac{\psi_{sn \ell}(r)}{r} \; Y^s_{m \ell} (\theta, \phi) \;  Z(y^1,y^2,y^3) \; e^{-i \omega t} \;.
\end{equation}
\noindent To determine the QNM behaviour, we have shown that we may use the Klein-Gordon equation. Recall that the Laplacian of a product space is the sum of its parts, such that
\begin{equation}
    \nabla^2 \Psi (\mathbf{z}) = \left( \nabla^2_{\rm BH} + \nabla^2_{\rm nil} \right) \Phi^{s}_{n \ell m}  (\mathbf{x}) Z(\mathbf{y})  \;.
\end{equation}
\noindent However, if we choose to impose a KK reduction, we may encode the higher-dimensional behaviour through an effective mass term representing a KK tower of states. This allows us to formulate the 7D scalar field evolution as a 4D ``massive" Klein-Gordon equation,
\begin{equation}
\frac{1}{\sqrt{-g}} \partial_{\mu} \left( \sqrt{-g} g^{\mu \nu} \partial_{\nu} \Psi \right) -\mu^2 \Psi= 0  \;,
\end{equation}
\noindent where
\begin{equation}
    \nabla^2_{\rm nil} Z (\mathbf{y}) = -\mu^2 Z(y^1,y^2,y^3) \;.
\end{equation}
\par Using the derivative of the tortoise coordinate $dr_*=dr/f(r),$  we extract the radial component of the QNM to produce a characteristic wave-like equation containing the QNF and the effective scalar potential,
\begin{equation} \label{eq:SWE}
    \frac{d^2\psi}{dr^2_*} + \left(\omega^2 - V(r) \right) \psi = 0 \;, 
\end{equation}
where
\begin{equation}
    V(r)=   \left(1-\frac{2M}{r} \right)\left(\frac{\ell (\ell+1)}{r^2}+\frac{2M}{r^3}+\mu^2 \right) \;.
\end{equation}

\par Within the QNM literature, a Klein-Gordon equation with a non-vanishing mass\footnote{It is worth noting that the origin and nature of this mass is rarely discussed in the context of QNMs.} has been used to describe the behaviour of massive 4D scalar QNMs in a black hole space-time \cite{SimoneWill1991_MassiveScalar,Ohashi2004_MassiveScalarQ,Konoplya2004_MassiveScalar,Dolan2007_MassiveScalarKerr,Dolan2011_MassiveVectorSchwarz,Konoplya2019,Seymour2020_MassiveScalar}. The reduction of our Schwarzschild-nilmanifold QNM equation to Eq. (\ref{eq:SWE}) allows us to draw upon known computational techniques and behaviours to constrain $\mu$. In the next section, we shall discuss the methods we employ here to compute the QNF spectrum from Eq. (\ref{eq:SWE}), after which we shall comment on the effect of $\mu$ on the QNFs and the implication thereof. 

\section{The QNF spectrum for the Schwarzschild-nilmanifold setup \label{sec:QNFspectrum}}

\subsection{Computing the QNFs}
\par There are several techniques established within the QNM literature that generate exact solutions for QNFs. These must contend with the technical challenges introduced by the inherently dissipative nature of the QNM problem. Since radiation is irrevocably lost at spatial infinity and at the event horizon, the system is not time-symmetric; the eigenvalue problem is consequentially non-Hermitian and the eigenvalues are complex. In general, the corresponding eigenfunctions are then not normalisable and do not form a complete set (see reviews \cite{refNollert1999,refBertiCardoso,refKonoplyaZhidenkoReview} for further discussion). To circumvent this problem, a method was developed in Refs. \cite{PoschlTellerMethod, refFerrMashh1, refFerrMashh2} that exploits the relationship between the QNMs of a potential barrier and the bound states of the inverted potential \cite{PoschlTellerMethod}, as explained in the introduction. The procedure involves fitting the effective QNM potential featured in Eq. (\ref{eq:SWE}) to a well-understood substitute (characterised by exponential decay and other key common features) for which analytic solutions are known. In the case of several black hole space-times\footnote{While the use of the inverted P{\"o}schl-Teller potential leads to the production of QNFs with errors $>1\%$ for Schwarzschild black holes with $\ell >2$, greater accuracy can be found for Schwarzschild-de Sitter and Reissner-Nordstr\"om-de Sitter black hole space-times, against which the P{\"o}schl-Teller potential exactly matches.}, the P{\"o}schl-Teller potential \cite{PoschlTellerPotential} can serve as the inverted effective potential and the QNF spectrum is extracted from the bound-state solutions.

\par However, physically-motivated numerical methods remain a popular alternative. For a spherically-symmetric black hole, QNMs can be treated as waves trapped on the photon sphere\footnote{The photon sphere of a non-rotating, spherically-symmetric black hole is comprised of circular null geodesics of fixed radius $r_c$. QNM behaviour can be compared with the photons orbiting this sphere: $\mathbb{R}e \{ \omega \}$ serves as the angular velocity while $\mathbb{I}m \{\omega \}$ refers to the instability timescale of the photon orbit.}, albeit gradually ``leaking out" \cite{refGoebel1972}. In Refs. \cite{refBHWKB0,refBHWKB0.5,refBHWKB1}, this scenario was interpreted as a scattering problem, where the effective QNM potential serves as a potential barrier that tends to constant values in the opposing asymptotic limits. From this framing, a modified WKB method was developed that exploited the Bohr-Sommerfeld quantisation condition of quantum mechanics to establish a semi-analytical technique to compute black hole QNFs.

\par The WKB formula involves the matching of asymptotic solutions across two turning points that are the roots of the effective QNM potential. With the aid of a Taylor expansion about the peak of the potential barrier $x=r_0$, it becomes possible to relate the ingoing and outgoing solutions of the wave-like Eq. (\ref{eq:SWE}) and thereby obtain an expression for the QNFs and their wave-function. At lowest order \cite{refBHWKB0}, this WKB method yields
\begin{equation} \label{eq:impreciseQNF}
    \omega^2 (\ell,n) \approx V(r_0) -i(n+1/2) \sqrt{-2V^{\prime \prime}(r_0)} \;,
\end{equation}
\noindent where derivatives with respect to $r$ are denoted by primes and $r_0$ represents the peak of the potential. From this simple expression alone, the dominant QNMs for a $s=2$ perturbing field may be computed with an accuracy of 6\% \cite{refBHWKB0.5}; at third-order \cite{refBHWKB1}, the accuracy improves to fractions of a percent \cite{Konoplya2019}. While the WKB method is far more successful than we would expect \cite{refFromen}, it is understood that this method produces more accurate results for QNFs when $\ell \gtrsim 2$ at lower orders \cite{Dolan2020_MassKerr}. However, even at higher orders (i.e. see the 12th-order WKB method established in Refs. \cite{Matyjasek2017,Matyjasek2019}), the method still works best for $\ell > n$, with further accuracy found at higher multipolar values. For low values of $n$, Eq. (\ref{eq:impreciseQNF}) demonstrates that the QNF can be closely determined by the height of its associated potential barrier, as well as its second derivative. 

\par There are, however, known limitations to the use of this modified WKB method: as reviewed in Ref. \cite{Konoplya2019}, care must be taken when applying the technique to instability analyses and contexts with large overtones, effective potentials with non-constant asymptotics, space-times with higher dimensions, QNMs of massive perturbing fields, etc. Specifically, in the case of massive scalar fields $-$ the QNM context which aligns most closely with our setup here $-$ the $\mu^2$ term in the effective potential produces an additional turning point beyond the two over which the WKB matching is traditionally applied. This becomes significant for large values of $\mu$, as the local minimum is lost \cite{Ohashi2004_MassiveScalarQ}. Physically, at a sufficiently large mass, the fields approach the \textit{quasiresonance regime}, at which point the the amplitudes in the asymptotic regions approach zero and the application of the WKB method is no longer feasible; damping becomes minimal, such that the modes become purely real and arbitrarily long-lived. 

\par To compute highly massive QNMs most accurately, one would have to take into account the minimum emergent on the right side of the peak and the consequent backscattering from that barrier. However, this is not strictly necessary provided the peak lies above the value to which the effective potential asymptotes i.e. $\mu^2 \leq V(r_0)$ (see section VI B of Ref. \cite{Konoplya2019} for explicit comparisons).  

\begin{table*}[t]
\begin{center}
\begin{minipage}{8.5cm}
\caption{\textit{Spin-0 QNFs for $n=0$ and $\ell=2$ for $0.0 \leq \mu \leq 0.7$ using the WKB at $\mathcal{O}(V^6)$, the P\"oschl-Teller (PT) method, and the Dolan-Ottewill (DO) expansion at $\mathcal{O}(L^{-6})$. }}\label{table:QNFs}%
\begin{tabular}{@{}CCCC@{}}
\hline\noalign{\smallskip}
\mu & \omega \; (DO) & \omega \; (WKB) & \omega \; (PT) \\
\noalign{\smallskip}\hline\noalign{\smallskip}
 0.0 & 0.4836- 0.0968i & 0.4836-0.0968 i & 0.4874-0.0979 i\\
 0.1 & 0.4868 - 0.0957i & 0.4868-0.0957 i & 0.4909-0.0968 i \\
 0.2 & 0.4963 - 0.0924i & 0.4963-0.0924 i & 0.5015-0.0936 i \\
 0.3 & 0.5124 - 0.0868i & 0.5123-0.0868 i & 0.5192-0.0881 i \\
 0.4 & 0.5352 - 0.0787i & 0.5351-0.0787 i & 0.5443-0.0800 i \\
 0.5 & 0.5653 - 0.0676i & 0.5649-0.0676 i & 0.5770-0.0690 i \\
 0.6 & 0.6032 - 0.0532i & 0.6022-0.0528 i & 0.6181-0.0541 i \\
 0.7 & 0.6500 - 0.0343i & 0.1396+0.2763 i & 0.6695-0.0312 i  \\
\hline\noalign{\smallskip}
\end{tabular}
\end{minipage}
\end{center}
\end{table*}

\par Recently, a numerical method was put forth in Ref. \cite{refDolanOttewill2009} that returns to the intuitive picture proposed by Goebel \cite{refGoebel1972}. Using a novel ansatz for Eq. (\ref{eq:SWE}) derived from the equations of motion for a test particle following the null geodesic of a spherically-symmetric black hole, Dolan and Ottewill iteratively construct a series expansion in inverse powers of $L=\ell+1/2$ for the QNF, \begin{equation}
    \omega = \sum^{\infty}_{k=-1} \omega_k L^{-k} \;.
\end{equation}

\par We have studied this technique extensively within the eikonal limit in Ref. \cite{refOurLargeL}; here, we find that the QNF emerges as a function of both $L$ and $\mu$ when the method of Ref. \cite{refDolanOttewill2009} is directly applied to Eq. (\ref{eq:SWE}):

\begin{gather} \nonumber
    \begin{aligned}
    &\omega(L,\mu)  = +\frac{1}{3}L -\frac{i}{6}L^0 +\Bigg[\frac{3 \mu^2}{2}+\frac{7}{648} \Bigg] L^{-1} \\
   & + \Bigg[\frac{5 i \mu^2}{4}-\frac{137 i}{23328} \Bigg] L^{-2}  + \Bigg[\frac{9 \mu^4}{8}-\frac{379 \mu^2}{432}
+\frac{2615}{3779136} \Bigg] L^{-3} \\
&+ \Bigg[ \frac{27 i \mu^4}{16}-\frac{2677 i \mu^2}{5184}+\frac{590983 i}{1088391168} \Bigg] L^{-4} \nonumber
    \end{aligned}
\end{gather}

\begin{gather}
    \begin{aligned}
      & +\Bigg[\frac{63 \mu^6}{16}-\frac{427 \mu^4}{576}+\frac{362587 \mu^2}{1259712}-\frac{42573661}{117546246144} \Bigg] L^{-5} \nonumber \\
& + \Bigg[ \frac{333 i \mu^6}{32}+\frac{6563 i \mu^4}{6912}+\frac{100404965 i \mu^2}{725594112} \\
     & +\frac{11084613257 i}{25389989167104} \Bigg] L^{-6} \;.\label{eq:QNFdo}
    \end{aligned}
\end{gather}

\par Our results are summarised in Table \ref{table:QNFs}, for which we set $\ell =2$ and $n=0$ to correspond to the least-damped/longest-lived ``fundamental mode" that dominates the QNM signal \cite{refBertiCardoso,refKonoplyaZhidenkoReview,Carullo2019_pyRing1}. We consider backscattering to be negligible. We find that the sixth-order WKB and the Dolan-Ottewill methods are in close agreement. We can ascribe the deviations in the P\"oschl-Teller results to the method's stronger reliance on the potential shape, where the P\"oschl-Teller potential is known to match closest to the inverted potential corresponding to a Schwarzschild black hole space-time with a positive cosmological constant \cite{refZhidenko2004}.

\subsection{The effect of a mass-like term on the QNF spectrum}

\par
\begin{figure}[h!]
%\begin{center}
\includegraphics[width=\linewidth]{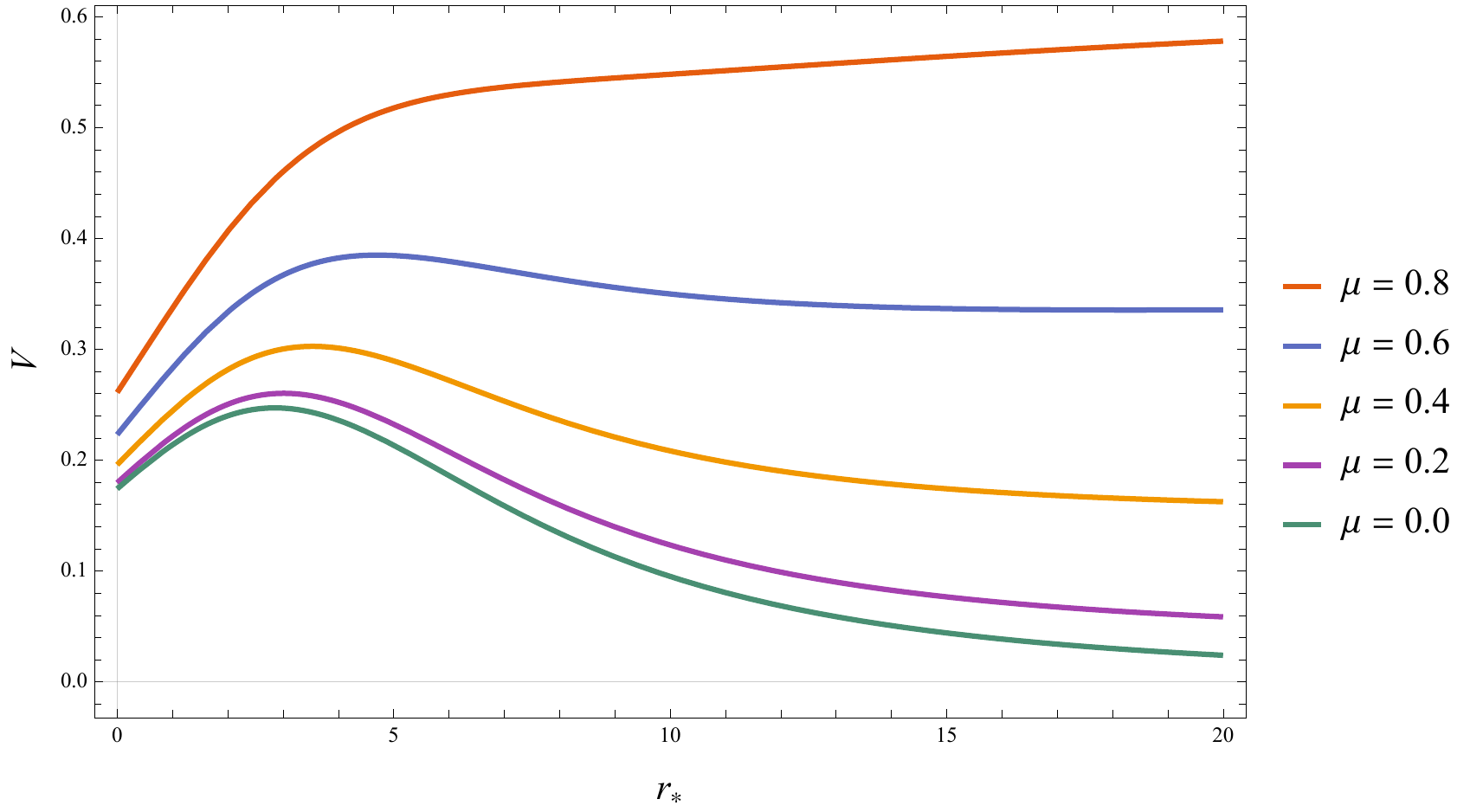}
\caption{\textit{The ($n,\ell)=(0,2)$ mode of the scalar potential of Eq. (\ref{eq:SWE}) for increasing values of the parameter $\mu$. Note that for $\mu = 0$, $V \rightarrow 0$ as $r_{*} \rightarrow \infty$ and the effective potential has a distinct peak. For $\mu \neq 0$, $V \rightarrow \mu^2$ as $r_* \rightarrow +\infty$. When $\mu^2 \gtrsim V(r_0)$, the peak is smoothed and the potential barrier is transformed into a potential step. }} \label{fig:potential}
%\end{center}
\end{figure}

 From the radial wave equation Eq. (\ref{eq:SWE}), the characteristic nature of the field is enclosed in the potential; from Eq. (\ref{eq:impreciseQNF}), we observe that the QNF value is strongly influenced by the potential. As such, it is useful to study the QNF spectrum in conjunction with Fig. \ref{fig:potential} to understand the effect of the mass-like term. We observe that $\mu$ elevates the potential: as $r_*$ increases, the potential no longer asymptotes to zero but instead approaches $\mu^2$. Beyond $\mu\approx 0.6$, the peak is smoothed out, suppressing the potential barrier and removing the local maximum.
\par From Table \ref{table:QNFs} demonstrating the fundamental QNM mode, we observe that $\mathbb{R}e \{ \omega \}$ increases steadily with $\mu$ whereas $\mathbb{I}m \{ \omega \}$ decreases. As $\mu$ approaches $0.7$, there is a discernible change in the QNF behaviour: a large jump in both the real and imaginary parts is observed for all three methods, with a pronounced difference in the WKB result for $\mu=0.7$: a sudden drop in $\mathbb{R}e \{\omega \}$ and shift from negative to positive in $\mathbb{I}m \{\omega \}$. This represents a breakdown in the method: while there is a known increase in the relative error for $\mu=0.7$ \cite{Konoplya2019}, we observe explicitly from Fig. \ref{fig:potential} that $\mu^2 > V(r_0)$ when $\mu=0.8$, which means the use of the WKB method is no longer appropriate. 

\par
\begin{figure}[t]
%\centering
\includegraphics[width=\linewidth]{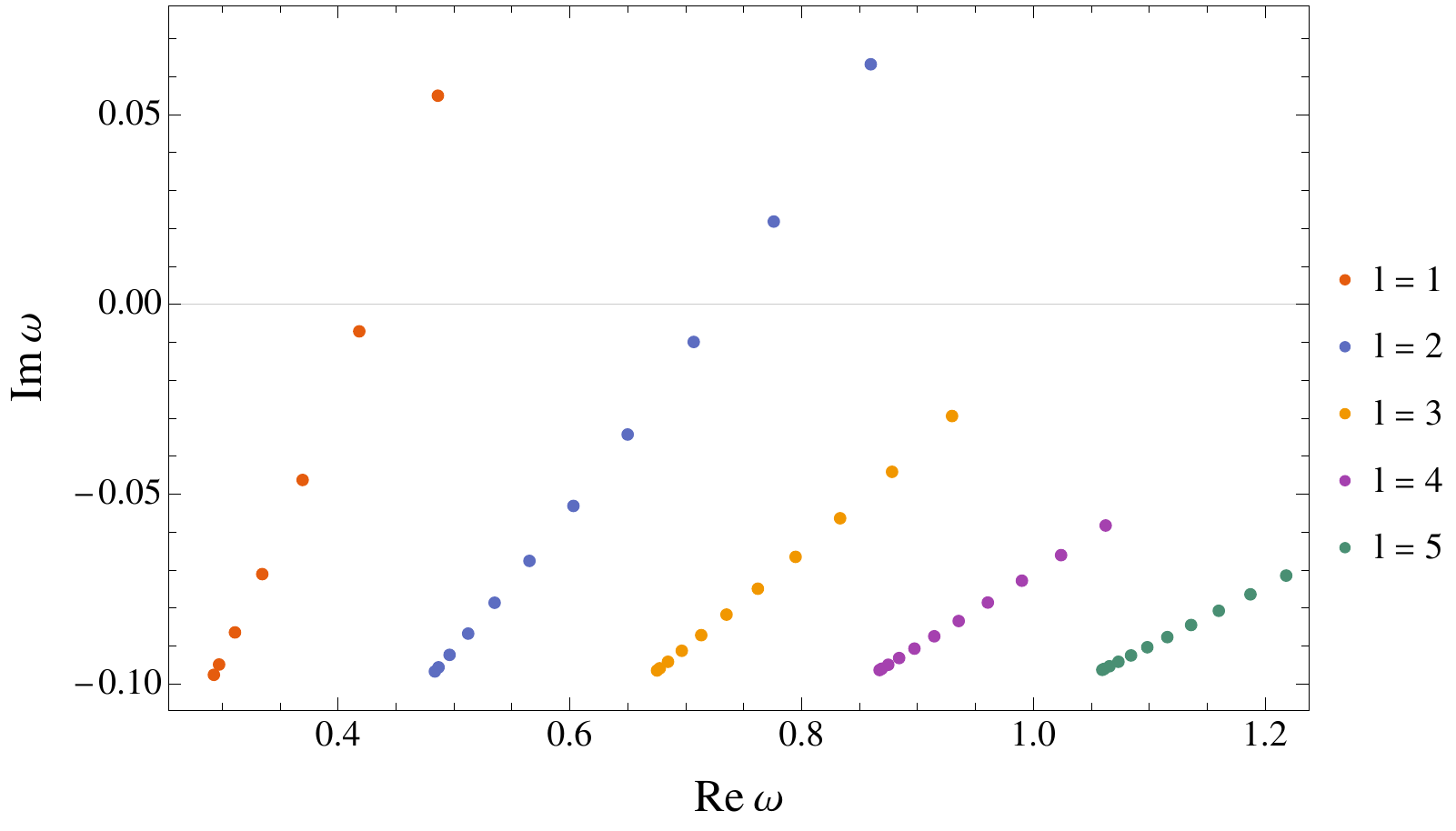}
\caption{\textit{The QNF spectrum of the extra-dimensional scalar field whose higher-dimensional contribution emerges as a mass-like term. We use the Dolan-Ottewill method to plot the imaginary components against the real for $\mu \in \{0,1\}$. Note that even for these small multipolar numbers, the range of the QNF decreases for increasing $\ell$.}}
\label{fig:QNFmu}
%\vskip 0.5cm
%  \caption{Imaginary part of the QNF plotted against the perturbing field mass.}
%  \includegraphics[width=9cm]{ImQNFmu.pdf}
%  \label{fig:ImQNF}
%\vskip 0.5cm 
%  \includegraphics[width=9cm]{ReQNFmu.pdf}
%      \caption{\textit{Real part of the QNF plotted against the perturbing field mass.}}
%  \label{fig:ReQNF}
\end{figure}

 However, there is also the physical interpretation to consider. In the geodesic picture, we understand that the flattening of the potential forbids the quantum tunnelling that allows the waves to ``leak out" from the system. In Ref. \cite{Dolan2020_MassKerr}, massive QNMs for which $\mathbb{R}e \{ \omega^2 \} > \mu^2$ are defined as ``propagative" and behave similarly to their massless counterparts, whereas $\mathbb{R}e \{ \omega^2 \} < \mu^2$ are ``evanescent" and contribute negligibly to the QNM spectrum for a perturbed black hole. This shift from propagative to evanescent is characterised by a change in sign in the imaginary part, as observed in Fig. \ref{fig:QNFmu}. As $\mu$ increases, the QNMs transition from propagative to evanescent; as the imaginary part goes to zero, the QNMs enter the quasiresonance regime \cite{Ohashi2004_MassiveScalarQ}, where the QNMs are arbitrarily long-lived. In this regime, the ingoing wave amplitude at the event horizon of the black hole is considered much smaller than the amplitude far from the black hole; since energy no longer ``leaks" from the system at spatial infinity, the QNMs behave as standing waves \cite{Konoplya2004_MassiveScalar}. 

\par While we focus on the $\ell=2$ mode that dominates the observed QNF spectrum \cite{Berti2005_BHspectroscopy,Carullo2019_pyRing1}, we can see in Fig. \ref{fig:QNFmu} that the oscillation timescale increases with the angular momentum number. This corresponds well to classical and quantum systems with which we are familiar, where the frequency of an oscillating wave increases with energy. Note, however, that as $\ell$ increases, the influence of $\mu$ wanes: the range of the QNF values converge to their massless counterpart for larger multipolar numbers.

\par In our extra-dimensional setup, the $\mu$ parameter serves as a manifestation of the extra dimensions, representing the KK tower of states. The analysis of the QNM potential and corresponding QNF spectrum conducted here demonstrates that only the ``propagative" QNMs can be used as a probe in extra-dimensional searches. This places an upper bound on $\mu$, such that $\mathbb{R}e \{ \omega^2 \} > \mu^2$. For a scalar test field in the Schwarzschild black hole space-time, we therefore consider the bound from the numerical analysis to be $\mu \lesssim 0.6.$

\subsection{An interpretation of $\mu$ in the QNM context \label{sec:Dimitrios}}

\par Within the QNM literature, potentials of the form provided in Eq. (\ref{eq:SWE}) have been approached primarily as a numerical problem (see e.g. Refs. \cite{SimoneWill1991_MassiveScalar,Ohashi2004_MassiveScalarQ,Konoplya2004_MassiveScalar}). However, a study of the $\mu$ parameter can also offer insights into the QNM problem at hand. To illustrate the role played by $\mu$ in QNM studies, and for a sense of the scales probed by QNFs influenced by this term, we discuss some examples from the literature. 

\par Physically, we understand $\mu$ to be of dimensions of inverse length, such that $m=\mu \hbar$ (under units of $G=c=1$). The corresponding Compton wavelength $\lambda_C=h/(mc)$ can then be related to the mass in eV using
\begin{equation}
    \lambda_C \times m = 1.24 \times 10^{-9} \;.
\end{equation}
\noindent For Compton wavelengths corresponding to astrophysical black holes $m_{\mathrm{BH}} \sim 10 M_{\odot}$, $\mu$ will correspond to very light particles of mass $m \sim 10^{-10}$ eV/c$^2$ \cite{Berti2014_BHprimer,Lagos2020_Anomalous}. 

\par Motivated by the long-lived nature of massive scalar QNFs, an investigation into the gravitational perturbations
coupled to the massive Klein-Gordon equation within a Schwarzschild space-time found similar masses, $m \sim 10^{-11}-10^{-12}$ eV/c$^2$ \cite{Herdeiro2014_MassiveFieldsBHs}. 

\par Of particular importance is the role played by the dimensionless parameter $M\mu$, where $M$ is the black hole Arnowitt-Deser-Misner (ADM) mass and $m  = \mu \hbar$ as before is the bosonic field mass. In the case of spinning black holes, this dimensionless parameter acts as a scaling for the suppression of the instability  timescale: when the Compton wavelength of the perturbing field is of the order of the black hole's radius, the dimensionless parameter scales as $M\mu \sim 1$, leading to the strongest super-radiant instabilities for the Kerr black hole \cite{refBertiCardoso,refKonoplyaZhidenkoReview}. This scenario is applicable also to light primordial black holes \cite{Sasaki2018_PBHreview}. 

\par In a study of Proca field QNMs in the Kerr space-time \cite{Dolan2020_MassKerr}, Dolan and Percival found $M\mu$ to be exceedingly large in the case of SM vector bosons, and extremely small for the photon with $\mu \lesssim 10^{-27}$ eV/c$^2$. Furthermore, they found that in the SM case, only evanescent modes for QNFs with $\ell + 1/2 \lesssim \mathcal{O}(M\mu)$ were predicted. For these extremely light photons, the QNF spectrum was anticipated to replicate that of the electromagnetic field, albeit with one extra longitudinal polarisation matching the QNF spectrum of a scalar field with mass $\mu > 0$.

\par Since a number of BSM conjectures depend on the existence of light or even ultralight particles (e.g. light scalars of mass $10^{-32} \leq m \leq 10^{-10}$ eV as in the ``string axiverse" scenarios \cite{Arvanitaki2009_StringAxiverse}, dark or hidden photons, and other candidates \cite{PDG2022}), massive QNMs may be useful in complementary searches for a variety of exotic signatures. 

\begin{figure}[t] 
\begin{center}
\includegraphics[width=0.4\textwidth]{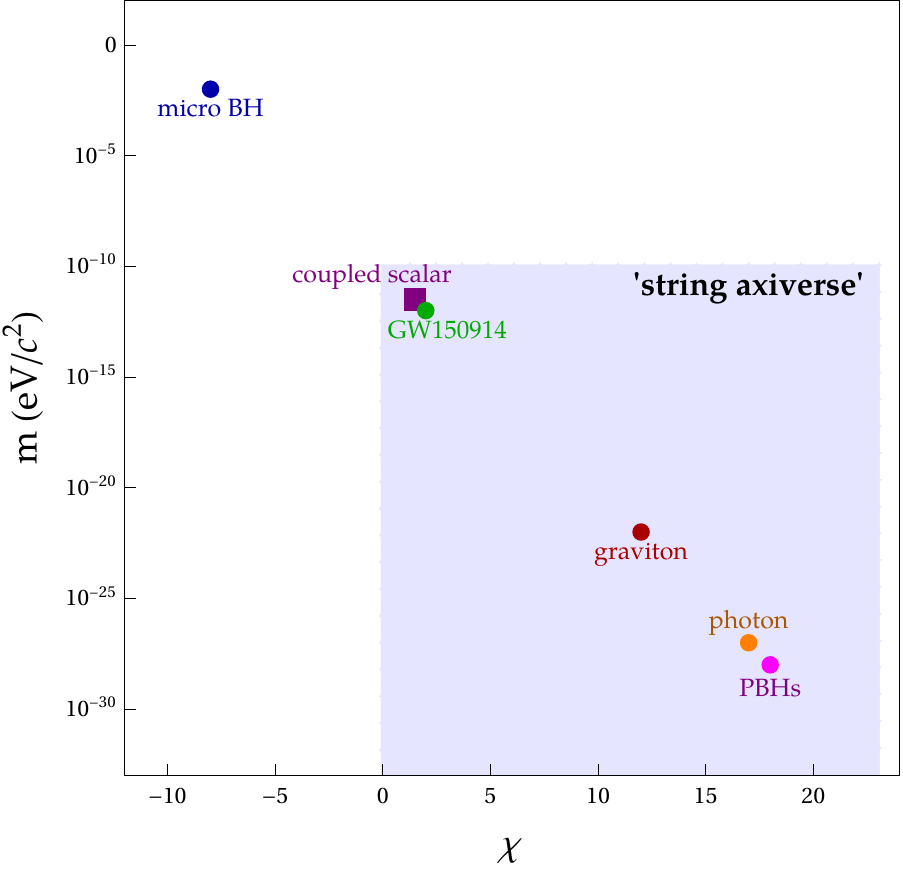}
\caption{\textit{For a sense of the scales probed by QNFs for $M\mu \sim 1$, we illustrate the magnitudes of $m$ and $\chi$ of Eqs. (\ref{eq:fieldmass}) and (\ref{eq:chi}) discussed in the text. We include the upper bounds for the photon mass \cite{Dolan2020_MassKerr} and the graviton \cite{LIGO2021_GWTC3-GRtest}, as well as the lower mass bound for massive primordial black holes \cite{Sasaki2018_PBHreview}. We note that particles of these magnitudes correspond to those of the ``string axiverse" scenarios \cite{Arvanitaki2009_StringAxiverse}.
}}\label{fig:chi}
\end{center}
\end{figure}

\par In our framework, we have positioned the $\mu$ parameter as an artefact of the extra-dimensional submanifold, representing the KK tower of states on the compact space. To obtain a sense of scale, we revert back to SI units such that the mass of the black hole and the $\mu$ parameter become
\begin{equation}
    M = \frac{G m^{\rm BH}}{c^2} \;\;\; \text{and} \;\;\; \mu = \frac{mc}{\hbar} \;. 
\end{equation}
\noindent From dimensional analysis, we can show that $M$ and $\mu$ have dimensions of length and inverse-length, respectively, such that $M\mu$ is indeed dimensionless. It is straightforward then that
\begin{eqnarray}
    M\mu &=& \frac{G m^{\rm BH} m}{\hbar c} \nonumber \\
    & \Rightarrow & m = \frac{1}{m^{\rm BH}} \frac{\hbar c}{G} M\mu \label{eq:fieldmass} \; .
\end{eqnarray}
\noindent With the values $\hbar c/G \sim 10^{-16}$ kg$^2$, $1 M_{\odot} \sim 10^{30}$ kg, and $M \mu \sim \mathcal{O} (1)$, we can scale the black hole mass as $m^{\rm BH} = 10^{\chi} M_{\odot}$ and thereby express the extra-dimensional contribution through 
\begin{equation} \label{eq:chi}
m \sim 10^{-\chi} 10^{-46} \text{kg}\sim 10^{-(\chi + 10)} {\text{eV/c}}^2 \;.
\end{equation}

\par We may use this expression to explore possible mass limits. From the well-known mass limit for non-\\
\noindent evaporating primordial black holes $m_{\mathrm{PBH}} \gtrsim 10^{15}$ g \cite{Sasaki2018_PBHreview}, $m \lesssim 10^{-28} \mathrm{ eV/c^2}$ such that $\chi \gtrsim 18$. On the other hand, $\chi \sim -8$ corresponds to a micro black hole of the same mass as the moon. For the 62 $\pm \; 4 \; M_{\odot}$ black hole remnant corresponding to the GW150914 event \cite{refLIGO}, $\chi \sim 2$. 

\par We can also contrast this against the dynamical lower bound on the graviton Compton wavelength $\lambda_g \geq 10^{13}$ km, as determined by the LVK collaboration at a 90$\%$ confidence (using null tests against the modified dispersion relation of massive-graviton theory introduced in Ref. \cite{YunesWill2012_LorenzViolations}). This in turn corresponds to the upper bound on the graviton mass $m_g \lesssim 10^{-22}$ eV/c$^2$ \cite{LIGO2016_TestingGRGW150914}, which leads to the bound $\chi \gtrsim 12$.  

\section{Constraints from GWs using QNMs \label{sec:GWs}}
\par In analogy to the electromagnetic waves produced by accelerating charges, GWs are generated by any massive body undergoing acceleration. This is a direct consequence of the relationship between mass and space-time curvature predicted by GR, where changes in the geometry occur corresponding to the movements of the massive body. Since gravity is weakly-interacting, the resultant ripples in space-time propagate throughout the universe unscreened. This property unlocks unique opportunities for studies into early-universe cosmology, since GWs decouple almost immediately after being produced and then propagate undisturbed throughout the universe; they may be the only way we can probe the time directly after the big bang \cite{Aggarwal2020_HighFGWs}. However, a consequence of this feeble nature of gravity is a severely limited collection of astrophysical events whose corresponding GW signatures lie within the sensitivity range of detectors. These can be classified into four possible GW sources: coalescing compact bodies, pulsars, supernovae (all of which are sources of \textit{deterministic} GWs), and a {\textit{cosmic GW background}} comprised of the stochastic GWs emergent in the wake of the big bang \cite{Caldwell2022_GWLHCSnowMass,Nature_GWreview}.    

\par The 90 GW events detected by the LVK collaboration \cite{refLIGO2018Run1,refLIGO2020Run2,refLIGO2021Run3a,refLIGOrecentRun3b} originate from the mergers of compact coalescing binaries, with binary black hole collisions remaining the most common. This is in part due to the energy output during a black hole collision (considered in Ref. \cite{refFerrari2008} as second only to the big bang), as well as the comprehensive understanding of the modelling of these two-body systems \cite{Jaranowski2005_GWanalysis}. Three distinct phases make up the gravitational waveform (where parentheses indicate the technique through which each phase is modelled),
\begin{itemize}
  \setlength\itemsep{0cm}
\item[(i)] inspiral: long, adiabatic stage as orbit shrinks and GW emission increases (post-Newtonian expansion);
\item[(ii)] merger: violent merger into a single black hole and GW emission peaks (numerical relativity simulations);
\item[(iii)] ringdown: final black hole emits damped GWs as it relaxes into a stationary state (black hole perturbation theory).
\end{itemize} 
\par Due to the weakly-interacting nature of GWs and the noise in which the signal is saturated, inferring the physical parameters of a GW source is a delicate process dependent on prior knowledge of the expected signal shape and the implementation of several \textit{a priori} assumptions (this is a highly non-trivial exercise, and we refer the interested reader to Ref. \cite{refLIGOguide} for details). However, the observations of GWs from these merger events allow for unique tests of GR within regimes previously beyond reach. 

\par In light of these regular GW detections and the promise of future GW observatories \cite{LIGO-India,LISA,CosmicExplorer}, interest in using GW data to constrain BSM models is building (see Ref. \cite{Caldwell2022_GWLHCSnowMass,Aggarwal2020_HighFGWs}). However, it is known that GW phenomenology is still in its infancy, unlike collider searches, where we have yet to obtain precise final state signatures for which we can search \cite{Yu2019}. This makes it difficult to constrain particle physics models with precision. Methods of searching for new physics predominantly rely on calculating the frequencies associated with symmetry breaking mechanisms to determine whether such signals lie within the sensitivity range of present or future GW detectors $-$ a strategy that long predates the detection of GWs  \cite{Will1997_BoundGravitonCBC,Hogan2000_GWsEDs1,Hogan2000_GWsEDs2}.
 Attempts to place bounds on the size and number of extra dimensions focus on the yet-undetected stochastic GW background rather than those emitted by compact coalescing bodies (see Ref. \cite{Yu2019}).
 
\par Within the GW community, searches for modified theories of gravity consider how GW signals may differ from those of GR in terms of their generation, propagation, and polarisation \cite{LIGO2019_GWTC1-GRtest,LIGO2020_GWTC2-GRtest_pyRing3,LIGO2021_GWTC3-GRtest}. In the case of massive gravity theories, for example, it is well understood that additional polarisation states must be considered to describe the extra degrees of freedom. While GR has only two tensor modes (i.e. plus and cross modes), a generalised metric theory of gravity can accommodate up to six polarisation modes: two tensor, two vector, and two scalar modes \cite{Eardley1973_GWsExtraPolarisations,deRham2017_GravitonMassBounds}. Similar effects can be seen in extra-dimensional setups e.g. Ref. \cite{AndriotGomez2017_EDsignGWs}; in such cases, however, these can often lie far beyond detectable range \cite{Cardoso2019_GWsEDsKK,Yu2019}. The situation is complicated further by the known difficulty in relating these null tests to one another \cite{Ghosh2022_RelatingGWtests}.   

\par For these reasons, we suggest a new avenue of pursuit by which to probe extra dimensions within extant GW data, that exploits the connection between QNM and GW studies. Inspired by tests for deviations from GR within the post-merger phase \cite{LIGO2020_GWTC2-GRtest_pyRing3,LIGO2021_GWTC3-GRtest}, we make use of one of the few tools dedicated to QNM analyses of GW data: the Python package \href{https://lscsoft.docs.ligo.org/pyring/}{\scshape{PyRing}} \cite{Carullo2019_pyRing1,refNoHair_pyRing2}. The package was recently developed to perform Bayesian parameter estimation, tests of GR, and other QNM analyses through a combination of observed GW data with simulation and numerically-generated waveform templates, following the Bayesian framework detailed in Ref. \cite{refLIGOguide}. Treating GR as the null hypothesis, {\scshape{PyRing}} tests for deviations from the QNF oscillation frequency ($\mathbb{R}e \{ \omega \}=\upomega$) and decay timescale ($1/\mathbb{I}m \{ \omega \} = \tau$):  
\begin{eqnarray}
\delta \upomega & = & \upomega^{\rm GR}(1 + \delta \upomega)  \; , \nonumber \\
\delta \tau & = & \tau^{\rm GR}(1 + \delta \tau) \;. \label{eq:dstructure}
\end{eqnarray}

\begin{figure}[t]
    \centering
    \includegraphics[width=\linewidth]{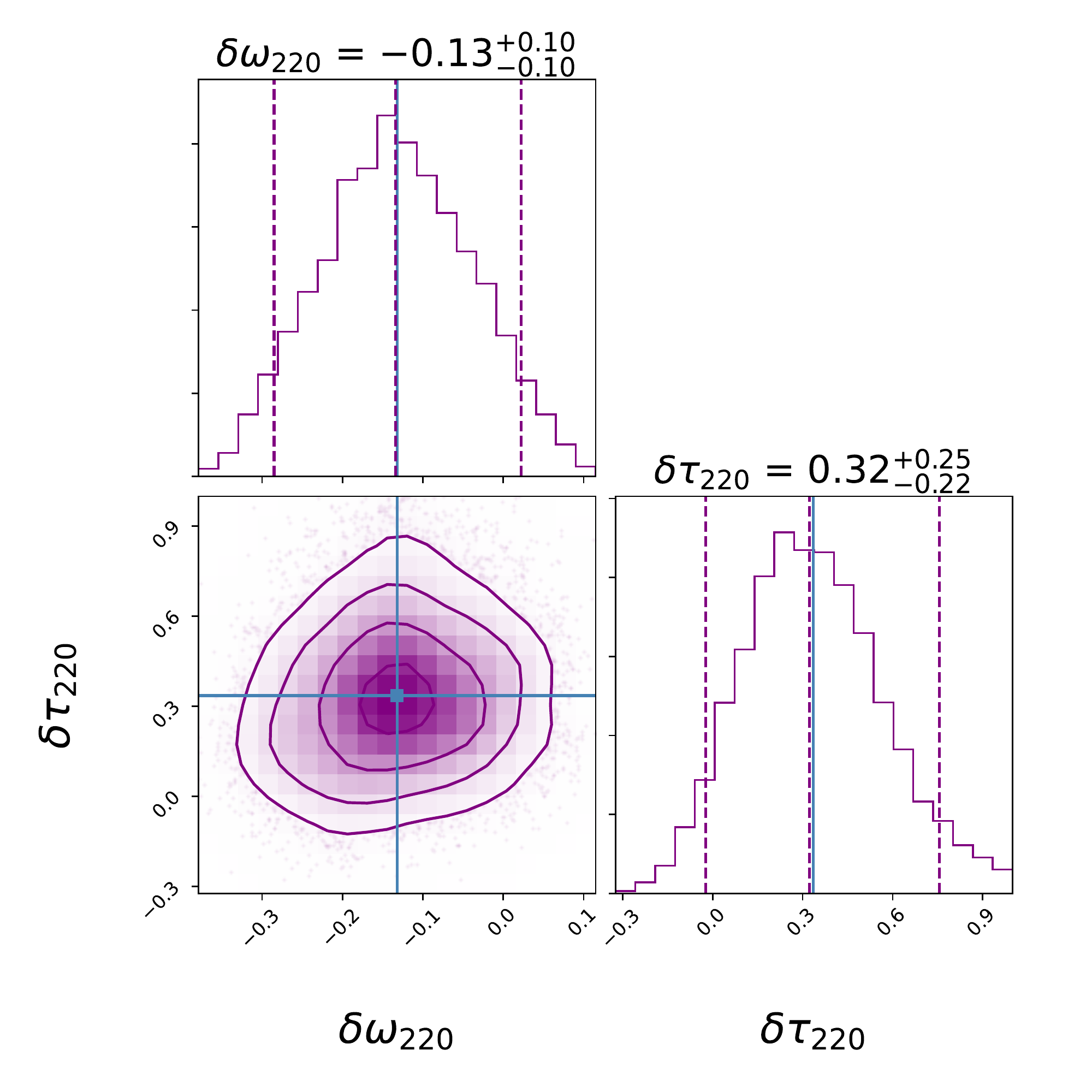}
    \vskip -0.3 cm
    \caption{\textit{As a proof-of-concept, we perform a rudimentary parameter estimation of the GR deviations using {\scshape{PyRing}} for event GW150914 (GW data sampled at 4096 Hz). We narrow priors to reduce computation cost. With \href{https://corner.readthedocs.io/en/latest/}{\scshape{Corner}}, we plot the 2D posteriors and 1D histograms on ($\delta \omega$, $\delta \tau$), where $(0,0)$ is the GR-predicted value. Dashed lines and contours demarcate the 90\% credible region; the blue line indicates the mean.
    \label{fig:plot}}}
\end{figure}

\begin{table}[t]
\caption{\textit{To correspond to the search for parametric deviations in GR, we structure our results for $\upomega =\mathbb{R} e \{ \omega \}$ and the damping time $\tau=1/\mathbb{I}m \{\omega \}$ as
$\upomega = \upomega^{\mu=0} \left(1 + \delta \upomega \right)$ and $ \tau = \tau^{\mu = 0} \left( 1 + \delta \tau \right)$, respectively. We use QNF results provided in column 2 of Table \ref{table:QNFs}. \label{table:param}}}
\begin{center}
\begin{tabular}{@{}CCCC@{}}
\hline\noalign{\smallskip}
\mu &  \omega (\ell,\mu)  &  \delta \upomega   &  \delta \tau \\
\noalign{\smallskip}\hline\noalign{\smallskip}
 0.0 & 0.4836 -0.0968i& 0.0000 & 0.0000 \\
 0.1 & 0.4868 -0.0968i& 0.0065 & 0.0113 \\
 0.2 & 0.4963 -0.0924i& 0.0262 & 0.0473 \\
 0.3 & 0.5124 -0.0868i& 0.0594 & 0.1149 \\
 0.4 & 0.5352 -0.0787i& 0.1066 & 0.2302 \\
 0.5 & 0.5653 -0.0676i& 0.1687 & 0.4306 \\
 0.6 & 0.6032 -0.0532i& 0.2472 & 0.8206 \\
\; 0.7 \; & \;\; 0.6500 -0.0343i \;\;& \;\; 0.3440 & 1.8181 \;\; \\
 \hline\noalign{\smallskip}
\end{tabular}
\end{center}
\end{table}
\noindent As a first exploratory step, we run this agnostic test of GR deviation in GW data from the GW150914 black hole merger event \cite{refLIGO} using the provided Kerr$_{220}$ waveform template corresponding to the $\ell=m=2$, $n=0$ mode (see Fig. \ref{fig:plot}). The analysis through {\scshape{PyRing}} is conducted in the time domain using publicly available data from the LVK collaboration \cite{LVK_OpenData}. To reduce computational cost, we employ medium-resolution data, simplified noise estimation, and simplified sampler settings, as well as tight priors. Specifically, we follow Ref. \cite{Carullo2019_pyRing1} in sampling 4096s of data from the Hanford and the Livingston LIGO detectors, sampled at 4096 Hz with the raw strain band-passed over $f \in [20,2028]$ Hz before being split into 2-second noise chunks. We set the trigtime in H1 to $t=126259462.423227$ s. We run the analysis over prior bounds for final mass $M_f~\in~[50.0,90.0]~M_{\odot}$, spin $a_f \in [0.6,0.9]$, amplitude $A_{220} \in [0.0,5.0 \times 10^{-20}]$, and phase $\phi_{220} \in [0,2 \pi]$. In testing for deviations from GR, we sample over $\delta \upomega, \delta \tau \in [-1,1]$.

\par To carry out its Bayesian inference, {\scshape{PyRing}} exploits the nested sampling algorithm of {\scshape{cpnest}} \cite{Veitch_NestedSampling,cpnest}. The package's implementation is based on an ensemble Markov chain Monte Carlo (MCMC) sampler, for which we only need to input the specifics of the analysis. We use 2048 live points and set the maximum MCMC steps to 2048, with the default 1234 seeds; at the end of the analysis, we are left with $\sim 8000$ independent samples. We visualise these results in Fig. \ref{fig:plot}.

\par Higher resolution data diminishes the impact of the time discretisation \cite{Cotesta2022_OvertonesRingdown}, while increased sampler settings lead to more precise results \cite{Veitch_NestedSampling,cpnest}.  For improved accuracy, we therefore make use of the hierarchical combination of LVK's strongest bounds on GR deviations to date \cite{LIGO2021_GWTC3-GRtest}:
\begin{eqnarray}
\delta \upomega_{220} & = & 0.02^{+0.07}_{-0.07} \; , \nonumber \\
\delta \tau_{220} & = & 0.13^{+0.21}_{-0.22} \label{eq:dstructureLIGO} \;.
\end{eqnarray}

\noindent To compare QNF computation with GW data, we consider our $\mu=0$ results to be equivalent to the GR prediction $(\delta \upomega,\delta \tau)=(0,0)$ i.e. $\upomega^{\mathrm{GR}} =\upomega^{\mu = 0}$. From the Dolan-Ottewill results  of Table \ref{table:QNFs}, we extract the parametric deviations to Table \ref{table:param}. We observe that the parametric deviations match the bounds predicted in Eq. (\ref{eq:dstructureLIGO}) for $\mu \sim 0.2$. If we exploit the QNF series expansion provided in Eq. (\ref{eq:QNFdo}), we can solve for $\mu$ explicitly. In doing so (for the real part and using the dominant $\ell =2,$ $n=0$ mode), we find that we can impose the upper bound
\begin{equation} \label{eq:mu_limits}
 \mu \lesssim 0.3681 \;.
\end{equation}
%"We find that it is harder to measure the departure of the mode decay times from their GR values than it is with the mode frequencies.\cite{Veitch2011_NoHairBayes}"
\noindent This serves as an upper bound on the sensitivity of QNFs to extra-dimensional KK resonances, as construct- ed in this framework. Using Eq. (\ref{eq:fieldmass}), we can explore the physical insights that can be extracted from this limit.

\par Since we have set $M=1$, we can interpret this as a bound on the dimensionless parameter $M\mu$. {\color{black}{As such, $M\mu \sim \mathcal{O}(0.1)$. Then for the final $M\sim 62M_{\odot}$ black hole remnant of GW150914, $\chi \sim 3$. This leads to the upper bound on the QNF probe,
\begin{equation}
    m \lesssim 10^{-13} \mathrm{ eV/c}^2 \;.
\end{equation}

\par In other words, we observe that applying static black hole QNFs as a direct probe into an agnostic extra-dimensional model demonstrates that QNFs cannot detect KK masses beyond roughly $m \sim 10^{-13} \mathrm{ eV/c}^2$.}} We note that particles of this mass correspond to light scalar hypotheses rather than the TeV-scale KK masses of typical extra-dimensional conjectures \cite{PDG2022}. 

%However, we have demonstrated that combining GW considerations with numerical QNF calculations does lead to more stringent limits on the BSM parameter space that can be explored.

\section{Conclusions}

\par In this work, we have considered a novel extra-dimen- sional setup comprised of a Schwarzschild black hole embedded in a 7D product space-time whose extra dimensions form a negative compact space $-$ specifically, a nilmanifold built from Heisenberg algebra. We have pursued a strategy for an extra-dimensional search using QNFs. By positioning the extra-dimensional contribution as an effective mass-like $\mu^2$ term in the QNM potential, we have demonstrated through a numerical study a possible upper bound on this $\mu$. For the scalar test-field and Schwarzschild space-time background considered here, $\mu \lesssim 0.6$.

\par Then, by using searches for parametric deviations from GR, we further constrain this probe to $\mu \lesssim 0.3681$. Via Eq. (\ref{eq:fieldmass}), we demonstrate that this corresponds to $m_{KK} \lesssim 10^{-13} \mathrm{ eV/c}^2$. The limit provided in Eq. (\ref{eq:mu_limits}) can therefore be interpreted as a detectability bound on the QNM probe into extra dimensions. In other words, with currently available signals, we find that KK masses higher than roughly $m_{KK} \sim 10^{-13} \mathrm{ eV/c^2}$ cannot be detected with QNMs.

\par However, there are a number of improvements that could be made to this preliminary study that may lead to more stringent bounds, particularly in the application to other BSM scenarios. For example, we would expect minor corrections from the use of the more astro- physically-relevant Kerr black hole space-time and gravitational QNFs; this would be necessary for greater precision than the order-of-magnitude study conducted in this work. More significantly, we recognise that this investigation was limited by the need to adopt an agnostic approach to our pursuit of evidence of extra dimensions. As the LVK collaboration develops more sophisticated and model-specific ringdown templates to test for parametric deviations in GR, it would be interesting to observe how theoretical frameworks can be adapted to the question of searches for extra-dimensional signatures in GWs.

\par A further open question is to what extent can we apply such constraints to place bounds on the size and number of extra dimensions. For example, a next step for this study could be to subject the mass spectrum of the toy dark matter model studied in Ref. \cite{Andriot2016_TowardsKK} to this result in order to extract tangible bounds on the radius of the nilmanifold extra dimensions herein constructed. Moreover, a detailed investigation of the propagation of GWs in nilmanifold spaces is reserved for a future work.

\par As acknowledged in Ref. \cite{LIGO2021_GWTC3-GRtest}, there has been substantial progress in GW research from the analytical, numerical, and experimental fronts. GW phenomenology and our ability to perform precision-level testing of GR, however, are still in their infancy. It is our hope that the simple setup we have provided here may be refined as our understanding of the applicability of GW detection in fundamental physics grows, bringing these tests to a new level of accuracy.

\begin{acknowledgements}

ASC is supported in part by the National Research Foundation (NRF) of South Africa; AC is supported by the NRF and Department of Science and Innovation through the SA-CERN programme and a Campus France scholarship; EL is supported by the French Government, via École Normale Supérieure de Lyon. AC extends her appreciation to the organisers and participants of the GWOSC Open Data Workshop 2022, which special thanks to those involved in the IP2I study hub hosted by the \textit{groupe Ondes gravitationnelles}.

\textbf{Software.}  LVK data are interfaced through
\texttt{GWpy}, with support through \texttt{LALSuite}. The publicly-available \texttt{pyRing} package can be found at: \url{https://git.ligo.org/lscsoft/pyring}. We use the
\texttt{cpnest: v0.11.4} \cite{cpnest} and \texttt{corner: v2.2.1} \cite{corner}. Other open-source python packages required by \texttt{pyRing} include \texttt{cython} \cite{cython}, \texttt{h5py} \cite{h5py}, \texttt{matplotlib} \cite{matplotlib}, \texttt{numpy} \cite{numpy}, \texttt{scipy} \cite{scipy}, and
\texttt{seaborn} \cite{seaborn}.
\end{acknowledgements}

\bibliographystyle{spphys}
\bibliography{EDfingerprints}% common bib file
%% if required, the content of .bbl file can be included here once bbl is generated
%%\input sn-article.bbl

%% Default %%
%%\input sn-sample-bib.tex%

\end{document}